\documentclass[aps,showpacs,floatfix,superscriptaddress,pre,11pt]{revtex4-1}

\usepackage{amsmath,amssymb}
\usepackage{verbatim,graphicx}

\usepackage{color}
\usepackage[all]{xy}
\newcommand{\Tr}{\rm Tr}
\newcommand{\eref}[1]{(\ref{#1})}
\newcommand{\nn}{\nonumber}
\newcommand{\be}{\begin{eqnarray}}
\newcommand{\ee}{\end{eqnarray}}

\newcommand{\bmat}{\left ( \begin{array}{cc} }
\newcommand{\emat}{\end{array} \right ) }

\def\Tr{\textrm{Tr}}

\newcommand{\beq}{\begin{equation}}
\newcommand{\beqs}{\begin{equation*}}
\newcommand{\eeq}{\end{equation}}
\newcommand{\eeqs}{\end{equation*}}

\allowdisplaybreaks

\begin{document}

\title{Spectral and thermodynamic properties of the Sachdev-Ye-Kitaev model}
\author{Antonio M. Garc\'\i a-Garc\'\i a}
\affiliation{TCM Group, Cavendish Laboratory, University of Cambridge, JJ Thomson Avenue, Cambridge, CB3 0HE, UK}
\email{amg73@cam.ac.uk}
\author{Jacobus J. M. Verbaarschot}
\affiliation{Department of Physics and Astronomy, Stony Brook University, Stony Brook, New York 11794, USA}
\email{jacobus.verbaarschot@stonybrook.edu}
\begin{abstract} 
  We study spectral and thermodynamic properties of the Sachdev-Ye-Kitaev model, a variant of the $k$-body embedded random ensembles studied for several decades in the context of nuclear physics and quantum chaos. We show analytically that the fourth
  and sixth order energy cumulants vanish in the limit of large number of particles $N \to \infty$ which is consistent with a Gaussian spectral density. However, for finite $N$, the tail of the average spectral density is well approximated by a  
  semi-circle law. The specific heat coefficient, determined numerically from the low temperature behavior of the partition function, 
is consistent with the value obtained by previous analytical calculations. For energy scales of the order of the mean level spacing we show that level statistics are well described by random matrix theory. Due to the underlying Clifford algebra of the model, the universality class of the spectral correlations depends on $N$. For larger energy separations we identify an energy scale that grows with $N$, reminiscent of the Thouless energy in mesoscopic physics, where deviations from random matrix theory are observed. Our results are a further confirmation that the Sachdev-Ye-Kitaev model is quantum chaotic for all time scales.
 According to recent claims in the literature, this is an expected feature in field theories with a gravity-dual. 
 
\end{abstract}


\maketitle
\section{Introduction}
The insurmountable technical difficulties posed by the theoretical description of the many-body nuclear forces have led to many effective descriptions of  nuclei to 
bypass the microscopic Hamiltonian.
A crude assumption is to replace the nuclear Hamiltonian by
a random matrix ensemble \cite{wigner1951,dyson1962a,dyson1962b,dyson1962c,dyson1962d,dyson1972,guhr1998} only constrained by global symmetries (the Wigner-Dyson ensembles). Surprisingly good agreement was found between spectral correlations of highly excited nuclei and the analytical predictions of random matrix theory for energy scales of the order of the mean level spacing.
Despite of its success, 
this approximation has evident shortcomings. The nuclear-shell model suggests that nuclear interactions are well described by a mean-field potential plus a residual two-body interaction while in the random matrix approach higher many-body interactions are equally important. Moreover it was noticed that the spectral density associated to these high energy nuclear excitations did not follow the semi-circle law, the random matrix theory prediction, but it is better approximated by the Bethe formula \cite{bethe1936}. 

In  response to these problems, a model of fermionic random $k-$body interactions of infinite range, the so called $k-$body embedded ensembles, was proposed more than forty years ago \cite{bohigas1971,bohigas1971a,french1970,french1971} as a more accurate stochastic description of nuclei. 
Although the interactions are random, 
the effective Hamiltonian is sparse and therefore deviations from the Wigner-Dyson ensembles were expected.
Indeed numerical \cite{bohigas1971a} and later analytical results \cite{mon1975} show that, in line with the experimental data, the spectral density is Gaussian for sufficiently small $k$, instead of following the semi-circle law. By contrast, spectral correlations are still close to the random-matrix prediction \cite{verbaarschot1984} for sufficiently close eigenvalues. For more information on the model, especially in the context of nuclear physics and quantum chaos, we refer to \cite{benet2003,gomez2011,brody1981,kota2014,kota2011a}.

Recently, similar models of fermions with $k-$body infinite-range interactions, called Sachdev-Ye-Kitaev models (SYK)  \cite{kitaev2015,maldacena2016,polchinski2016,engels2016,almheiri2015,magan2016,danshita2016,garcia-alvarez2016,bagrets2016,sachdev2015,you2016,Gross:2016kjj}, and originally introduced in the study of spin liquids \cite{sachdev1993}, are being intensively investigated in a completely different context: holographic dualities in string theory \cite{maldacena1998}. 
Based on the same pattern of conformal symmetry breaking, it has been speculated \cite{kitaev2015,maldacena2016,polchinski2016,engels2016,almheiri2015,jensen2016,cvetic2016} that, in the infrared limit, the holographic dual of an Anti-deSitter (AdS) background in two bulk dimensions AdS$_2$ is closely related to one of the variants of the SYK model, namely, a model of $N$ Majorana fermions \cite{kitaev2015} in zero spatial dimensions with random two body interactions of infinite range. Green's functions \cite{bagrets2016,jevicki2016,maldacena2016,polchinski2016}, thermodynamic properties \cite{sachdev2015}, such as the low temperature limit of the entropy, and also out of equilibrium features \cite{maldacena2016} such as the exponential growth of certain out-of-time-ordered correlators are strikingly similar in both models. The latter, related to quantum corrections in the gravity dual \cite{maldacena2015}, is also a signature of quantum chaotic features. More interestingly, it is believed that the SYK model may describe the low energy limit of a higher dimensional gauge theory with a string theory dual still to be named.
Very recent results \cite{witten2016} suggest that disorder is not strictly necessary for a gravity-dual interpretation.

Despite these advances, the description of many aspects of the SYK model dynamics still poses severe technical, both numerical and analytical, challenges.
In closely related problems such as quantum chaos and disordered systems, the spectrum and level statistics provide a rather comprehensive
description
of the quantum dynamics without the need of the more expensive computation of eigenvectors. 
In the context of the SYK model, spectral correlations
have so far been investigated in \cite{you2016}, where level repulsion was found,
typical of a disordered metal, though its strength changes with  the number of particles $N$ modulo 8. 

Here we aim to fill this gap by carrying out an extensive analysis of the spectral density, thermodynamic properties, and both short-range and long-range spectral correlations of the SYK model, with $N$ Majorana fermions.

Our main results are summarized as follows: we show analytically that in the $N \to \infty$ limit the fourth and sixth cumulant of the spectral density vanish which strongly suggests that it is Gaussian. However its tail at finite $N$, that controls the specific heat, is well approximated by the semi-circle law.
Results from exact diagonalization, for up to $N=36$ Majorana fermions, are fully consistent with the analytical findings, including results for the
entropy and the specific heat.
Spectral correlations that test short range correlation as the level spacing distribution are in good agreement with the random matrix prediction. We find that, in agreement with \cite{you2016}, the Bott periodicity of the Clifford algebra that governs the Majorana fermions labels the global symmetries of the model. However we have observed systematic deviations from the random matrix predictions, for sufficiently well separated eigenvalues, that suggest that the model is not ergodic for short times. The point of departure from the universal results of random matrix theory increases with $N$ which is a strong indication of the existence of a Thouless energy \cite{altshuler1988,braun1995,bertrand2016} for the system.

This paper is organized as follows: in the next section
we  introduce the model and discuss its spectral density. The thermodynamical
properties of the model are evaluated in section III. Spectral
correlations are computed in section IV. We finish with concluding
remarks and some ideas for future research in section V. Some technical details involving the calculation
of the cumulants and the symmetry properties of the gamma matrices
are worked out in two appendices.
 
\section{The spectral density}
 
Kitaev recently introduced \cite{kitaev2015} a model of interacting fermions aimed to explore its potential as a gravity-dual. The Hamiltonian is given by,

 \begin{equation}\label{hami}
 H \, = \, \frac{1}{4!} \sum_{i,j,k,l=1}^N J_{ijkl} \, \chi_i \, \chi_j \, \chi_k \, \chi_l \, ,
 \end{equation}
 where $\chi_i$ are Majorana fermions that verify 
 \begin{eqnarray}
  \{ \chi_i, \chi_j \} = \delta_{ij}.\label{clif}
 \end{eqnarray}

 The fermions are coupled by Gaussian distributed random variables  $J_{ijkl}$
 with probability distribution,
 \begin{equation}
 P(J_{ijkl}) \, = \, \sqrt{\frac{N^3}{12 \pi J^2}} \exp\left( - \, \frac{N^3J_{ijkl}^2}{12J^2} \right) \, .
 \end{equation}

 We note that Eq. (\ref{clif}) is the defining relation of an Euclidean $N$-dimensional Clifford algebra. Many interesting features of the model are a direct consequence of Clifford algebra properties. For instance, the Bott periodicity of the Clifford algebra suggests that the global symmetries of the Majorana fermions, that to some extent control the spectral properties of the model, are sensitive to the arithmetic nature of $N$. We shall see that this is indeed the case when we study level statistics later in the paper.  
 It will also be helpful for our first objective: to derive analytical results for
 the many-body spectral density.

 We will follow the strategy of Mon and French \cite{mon1975} of evaluating  moments of the spectral density. In this model, this is again facilitated by noticing that the Euclidean Clifford algebra in $N$ dimensions of the Majorana fermions Eq. (\ref{clif}) is shared by Euclidean Dirac $\gamma$ matrices. Therefore it is possible to employ the full machinery developed in that context to compute the trace of a large number of Majorana fermions, a key part in the calculation of energy moments. We leave the details of the calculation to  appendix B. Here we just define the moments, sketch the main steps of the calculation, and give the 
 final expression as a function of the number of particles $N$. Since the Gaussian disorder distribution is an even function, all odd moments will vanish. From now on we will focus only on the even ones:
 
 \begin{eqnarray}
 M_{2p}(N)= \langle \Tr H^{2p}\rangle \label{moment}
 \end{eqnarray}
 where $p = 1,2,3 \ldots$, $\langle \ldots \rangle$ stands for spectral and ensemble average.  
 The strategy to evaluate $M_p(N)$ is straightforward: we first
 perform the Gaussian average, equivalent to summing over
 all possible contractions according
 to Wick's theorem, and then we evaluate each of these terms, involving the trace of products of $\gamma$ matrices, by using properties of $\gamma$ matrices in $N$ Euclidean dimensions. 

 Denoting the product of four $\gamma$ matrices by $\Gamma_\alpha$, we have
 that
the moments are given by
 \be
 M_{2p} = \left \langle \Tr \sum \prod_{k=1}^{2p} J_{\alpha_k} \Gamma_{\alpha_k}
 \right \rangle.
\ee
 The Gaussian average over the random couplings $J_\alpha$ of the Hamiltonian (\ref{hami}), denoted by $\langle \cdots \rangle$,  is equal to the sum over all possible
contractions.
In the limit $N \gg 2p$ almost all  $\Gamma_\alpha$ have no
overlapping indices so that they commute. Because of
\be
\Gamma_\alpha^2 =1,
\label{gam2}
\ee
 we find that in this case all $(2p-1)!!$ contractions
give the same contribution resulting in the moments
\be
M_{2p} =(2p-1)!!\langle J_\alpha^2\rangle^p 2^{N/2}.
\label{semi-moments}
\ee
These are the moments of a Gaussian distribution resulting in a Gaussian spectral
   density. 
  We have evaluated the exact analytical result for $M_4$ and $M_6$. This requires the evaluation of diagrams that are subleading in $N$.
For that purpose it is helpful to note that when we have common $\gamma$ matrices in $\Gamma_\alpha$  and  $\Gamma_\beta$
   they commute or anti-commute depending on the number of common
   $\gamma$ matrices. This results in large cancellations suppressing the contribution 
   of intersecting diagrams. 
Following this procedure the first two non-trivial normalized cumulants, $\kappa_4$ and $\kappa_6$, are easily obtained as a function of $N$ from the moments $M_{2p}(N)$ (see Appendix B for details),

             \be \label{moment4}
          \kappa_4(N)&=&  -\frac{32(N-4)(N^2-11N+36)}{N(N-1)(N-2)(N-3)},
          \ee

with large $N$ asymptotics $-32/N$ and
     \be
     \kappa_6(N)= \frac
           {512 (N-4) (11 N^5-304 N^4+3535 N^3-21302 N^2+65856 N-82656)}{(N-3)^2 (N-2)^2 (N-1)^2 N^2},
          \label{moment6}
\ee
 with large $N$ asymptotics $512\times11/N^2$ where from now on we set $J=1$. 
 
 For higher moments the combinatorial problem becomes increasingly difficult and the final expressions
 are rather cumbersome. However these few cumulants already contain interesting information. 
  \begin{figure}[t!]%
  	\centering
  	\resizebox{0.48\textwidth}{!}{\includegraphics{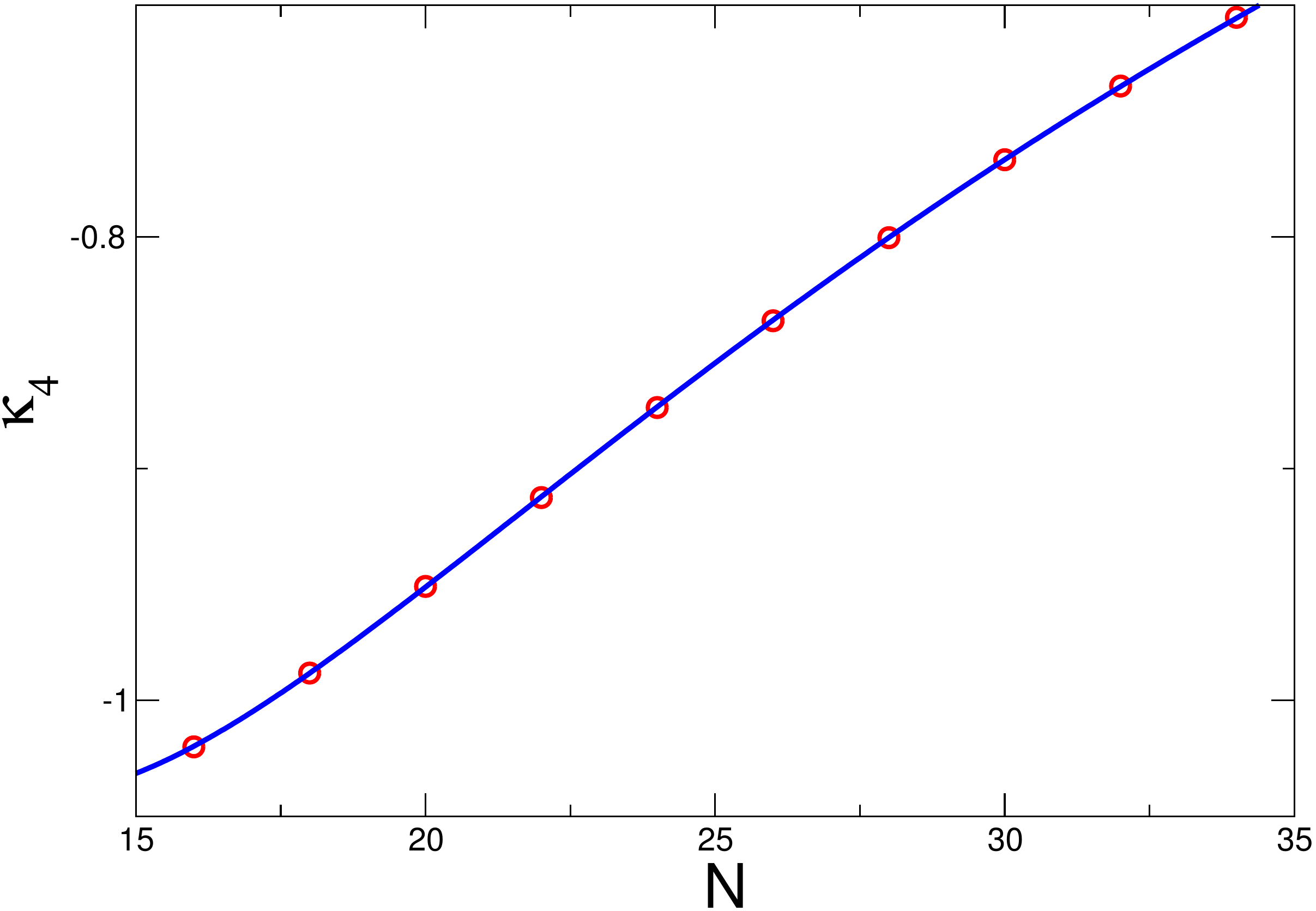}}
  	\resizebox{0.48\textwidth}{!}{\includegraphics{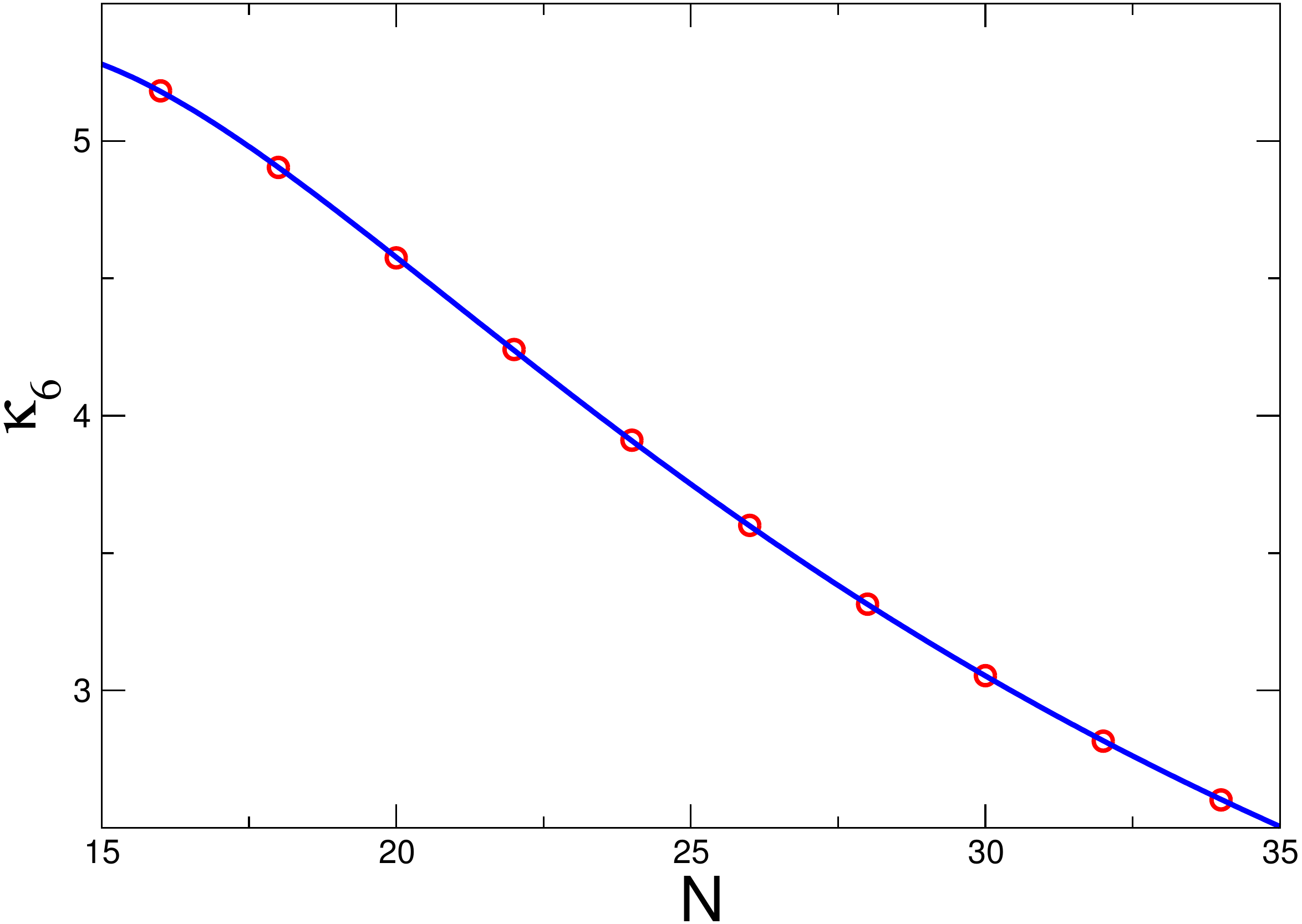}}
  	\caption{The fourth and sixth normalized energy cumulant related to the Hamiltonian (\ref{hami}) as a function of the system size $N$. The circles correspond to the numerical results obtained by exact diagonalization after spectral and ensemble average. At least a total of $5\times10^5$ eigenvalues were employed for each $N$. The solid line is the analytical prediction for the fourth (left) Eq.(\ref{moment4}) and sixth cumulant (right) Eq.(\ref{moment6}). }
  	\label{fig1}
  \end{figure}

   \begin{figure}[t]%
   	\centering
   	\resizebox{0.75\textwidth}{!}{\includegraphics{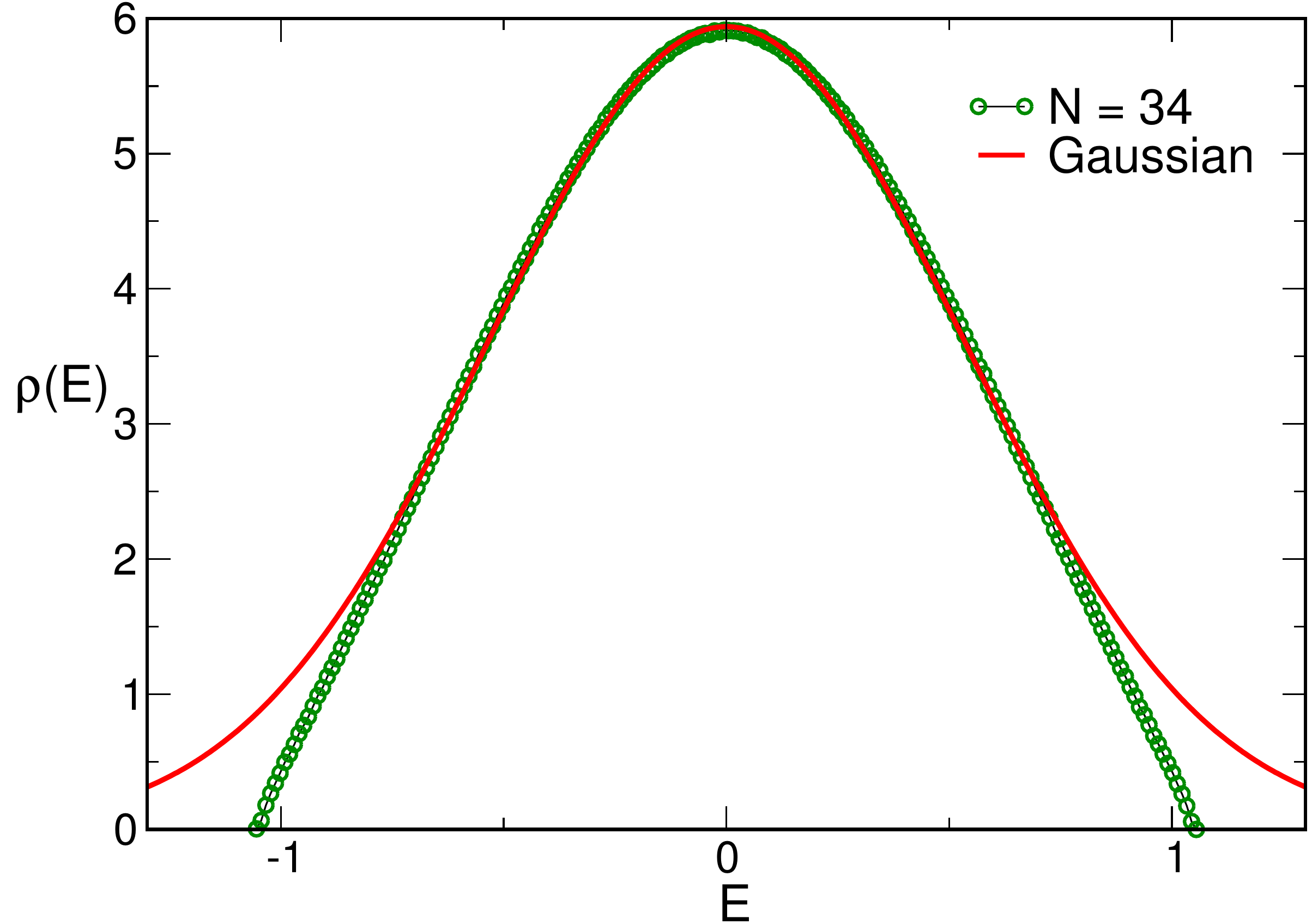}}
   	\caption{Spectral density $\rho(E)$ as a function of the energy $E$. The solid line is the analytical prediction valid in the $N \to \infty$ limit. Circles are the numerical spectral density for the largest size $N=34$ for which we can obtain all eigenvalues of the Hamiltonian. Except for the tails, the agreement with the numerical results is very good.}
   	\label{fig2}
   \end{figure}
As we have seen above, for $N \to \infty$ the normalized cumulants vanish for orders $8p \ll N$. This is a distinctive feature of a Gaussian distribution. Therefore the average analytical spectral density converges (non-uniformly)
 to a Gaussian of zero average and variance equal to
 $6/N^3$.
 
 We note that a Gaussian spectral density is expected for models with an entropy
 $S = N f(E/N)$ in the large $N$ limit. The only requirement is that $f$ is a smooth function that
 has a maximum. Gaussian behaviour in the central part of the spectrum, assuming a maximum at $E = 0$, results after expanding $f$ around the maximum.

 In Fig.~\ref{fig1}  we compare the analytical predictions
 Eqs. (\ref{moment4}-\ref{moment6}) of the normalized fourth and sixth
 cumulants with numerical results obtained by using exact diagonalization techniques. The agreement is excellent. 
  
 In Fig. \ref{fig2} we depict the  average spectral density for $N = 34$, the largest size for which we can obtain numerically the full spectrum, with the analytical prediction, a Gaussian distribution
with a variance that has been fitted to the data. Here the agreement is good but we observe clear deviations in the tail of the density. The reason for that discrepancy is that corrections to the Gaussian distribution, as described by the moments above,  are still of order one for $N = 34$. We were unable to compute analytically the leading $N$ corrections to the Gaussian density of states.
However, in the next section, we carry out a detailed numerical analysis of the tail of the average spectral density.
 
\begin{figure}[t!]%
	\centering
	\resizebox{0.85\textwidth}{!}{\includegraphics{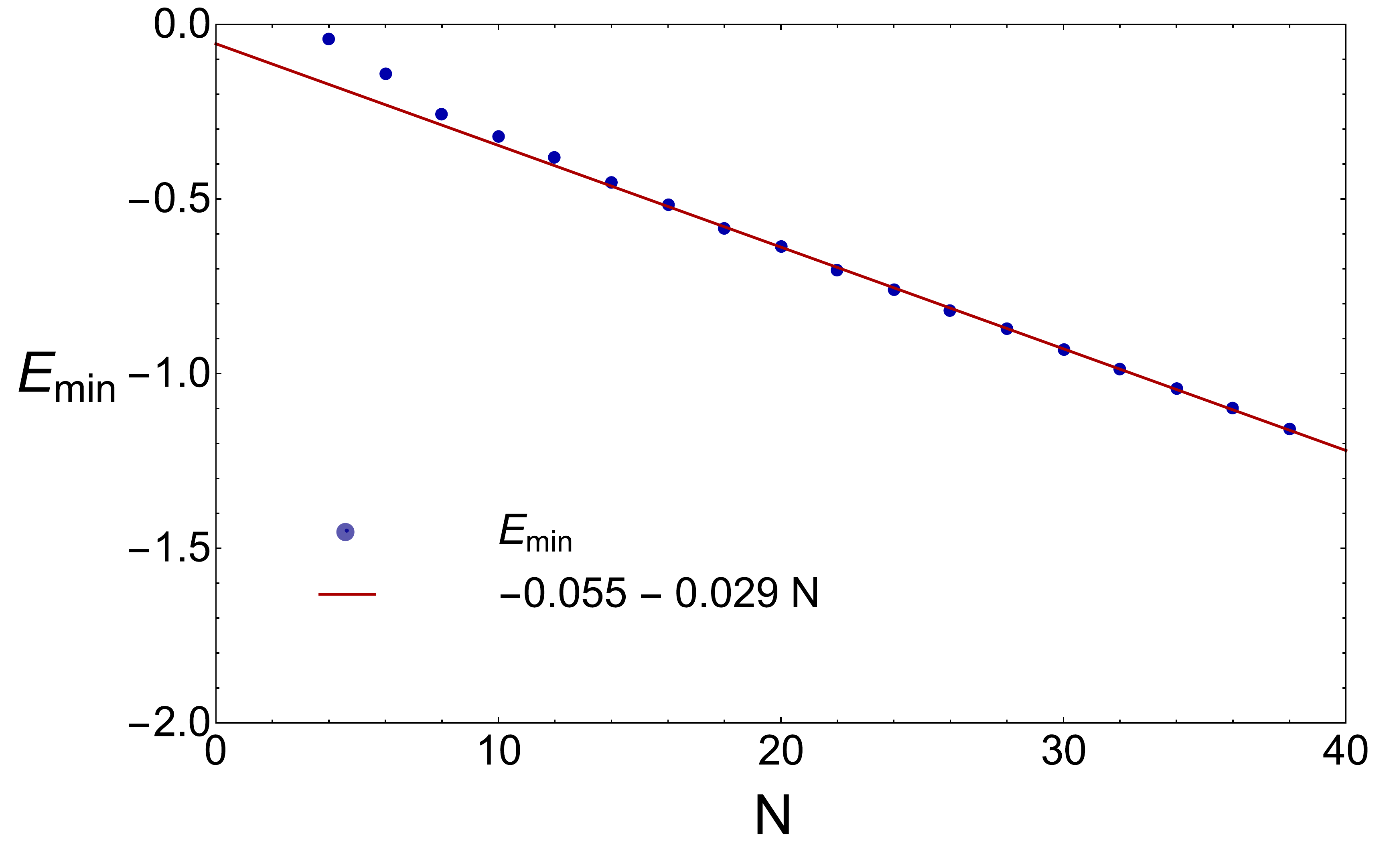}}
	\caption{Ensemble average of the smallest eigenvalue as a function of the system size $N$.
		For $N \gg 1$ we observe that it decreases linearly with $N$. This is an expected feature for
		a system of $N$ interacting fermions.}
	\label{fig4}
\end{figure}

 \section{Thermodynamic properties in the low temperature limit}
 Part of the renewed interest in the SYK model stems from the fact that its low temperature properties are similar to those of a 
 gravity background that in the infrared limit is well described by AdS$_2$ geometry. Typical features includes a finite entropy at zero temperature, a ground state energy that is extensive in the number of particles, and a specific heat linear in temperature but with a prefactor different from that of free fermions. There are already approximate analytical predictions \cite{maldacena2016,jevicki2016} in the literature
 for these observables. Exact numerical diagonalization of the SYK Hamiltonian Eq. \eref{hami} was employed in \cite{maldacena2016} to compute the zero temperature entropy \cite{maldacena2016}. We are not aware of exact diagonalisation results for the specific heat or the ground state energy. In this section we address this problem by a detailed numerical study of the tail of the spectrum that controls the thermodynamic properties in the low temperature limit. 
We start with the ground state energy. 
 The lowest eigenvalue of the SYK Hamiltonian, $E_{\rm min}$, is the ground state energy of the SYK model with $N$ Majorana fermions Eq. (\ref{hami}). 
 Due to the fermionic nature of model we expect $E_{\rm min}$ to be proportional to $N$. In Fig. \ref{fig4} we
 show the ensemble average of $E_{\rm min}$ versus $N$ and it indeed
 shows a nice linear asymptotic dependence on the dimension $N$.

From  a careful fitting of the numerical data we find that
 the tail of the spectrum 
 is well  approximated by
 \be\label{denfit}
\rho_{\rm tail}(E) =  2^{N/2} a (E-E_{\rm min})^{1/2}[1 + b (E-E_{\rm min})],
\ee
which also determines the low-temperature limit of the partition function,
\be
Z(\beta) &=& \int_{E_{\rm min}}^\infty e^{-\beta E} \rho(E),\nn\\
&\approx& \int_{E_{\rm min}}^\infty e^{-\beta E} \rho_{\rm tail}(E),\nn\\
&=& \frac {a\sqrt\pi } 2 \frac 1{\beta^{3/2}}e^{-\beta E_{\rm min}}
\left (1+ \frac 3{2\beta}  b \right). 
\label{zlev}
\ee
The low temperature limit of the SYK model is given by \cite{maldacena2016,jevicki2016},
\be \label{zmal}
Z(\beta) = \frac {c_0}{\beta^{3/2} }e^{-\beta E_0 + S_0 + \frac c{2\beta J}}, 
\ee
where the ground state energy, $E_0$, the entropy $S_0$ and the
specific heat coefficient, $c$, are all proportional to $N$. The prefactor $\beta^{-3/2}$ is an order one contribution coming from one-loop quantum corrections and $c_0$ is a temperature independent constant. 
Comparing to Eq. \eref{zlev} we can make the identification
\be\label{S0c}
E_0 &=& E_{\rm min},\\
S_0 &= &\frac N2 \log 2 + N \frac d{dN} \log a. \nonumber
\ee

\begin{figure}%
	\centering
	\resizebox{0.49\textwidth}{!}{\includegraphics{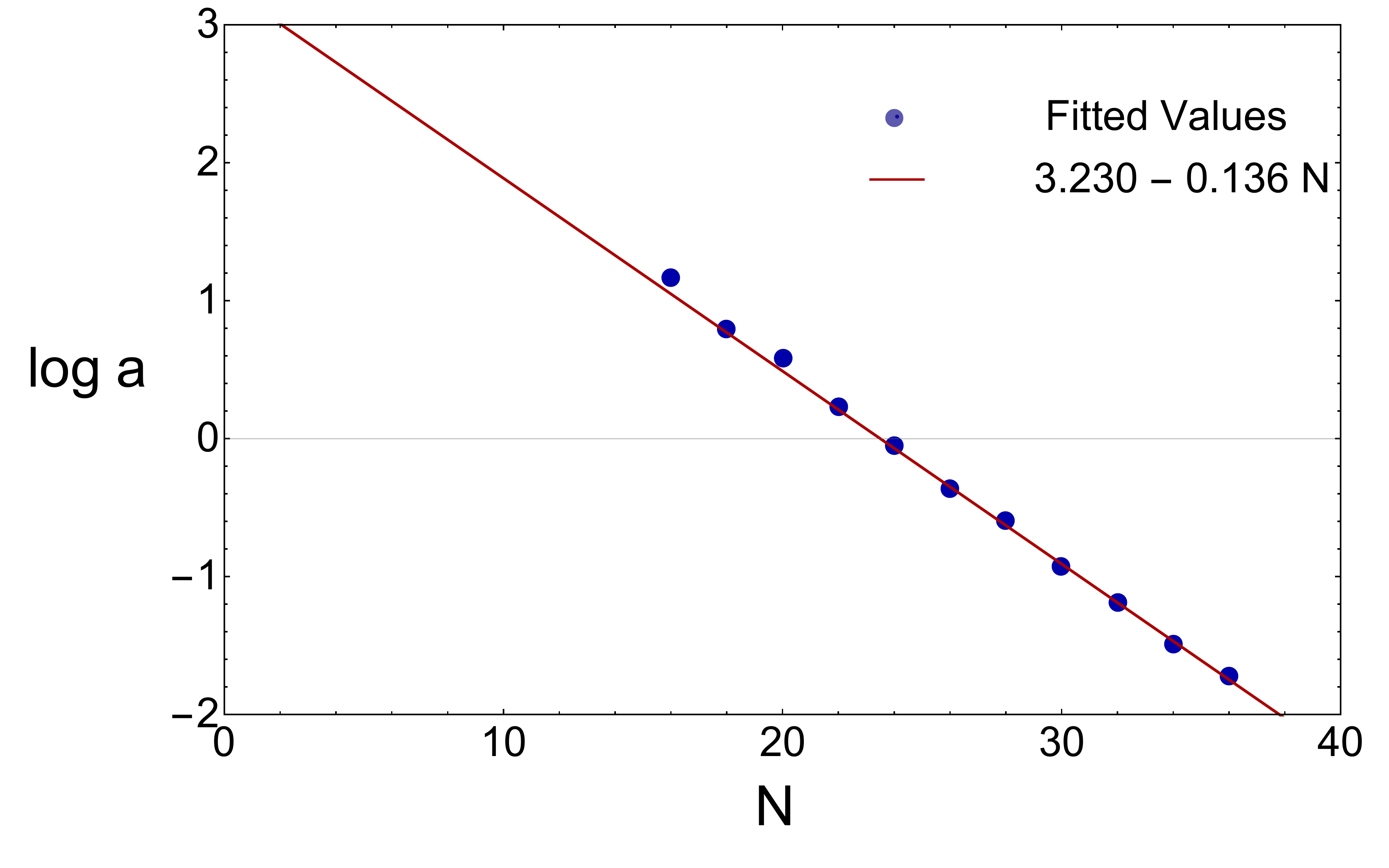}}
	\resizebox{0.49\textwidth}{!}{\includegraphics{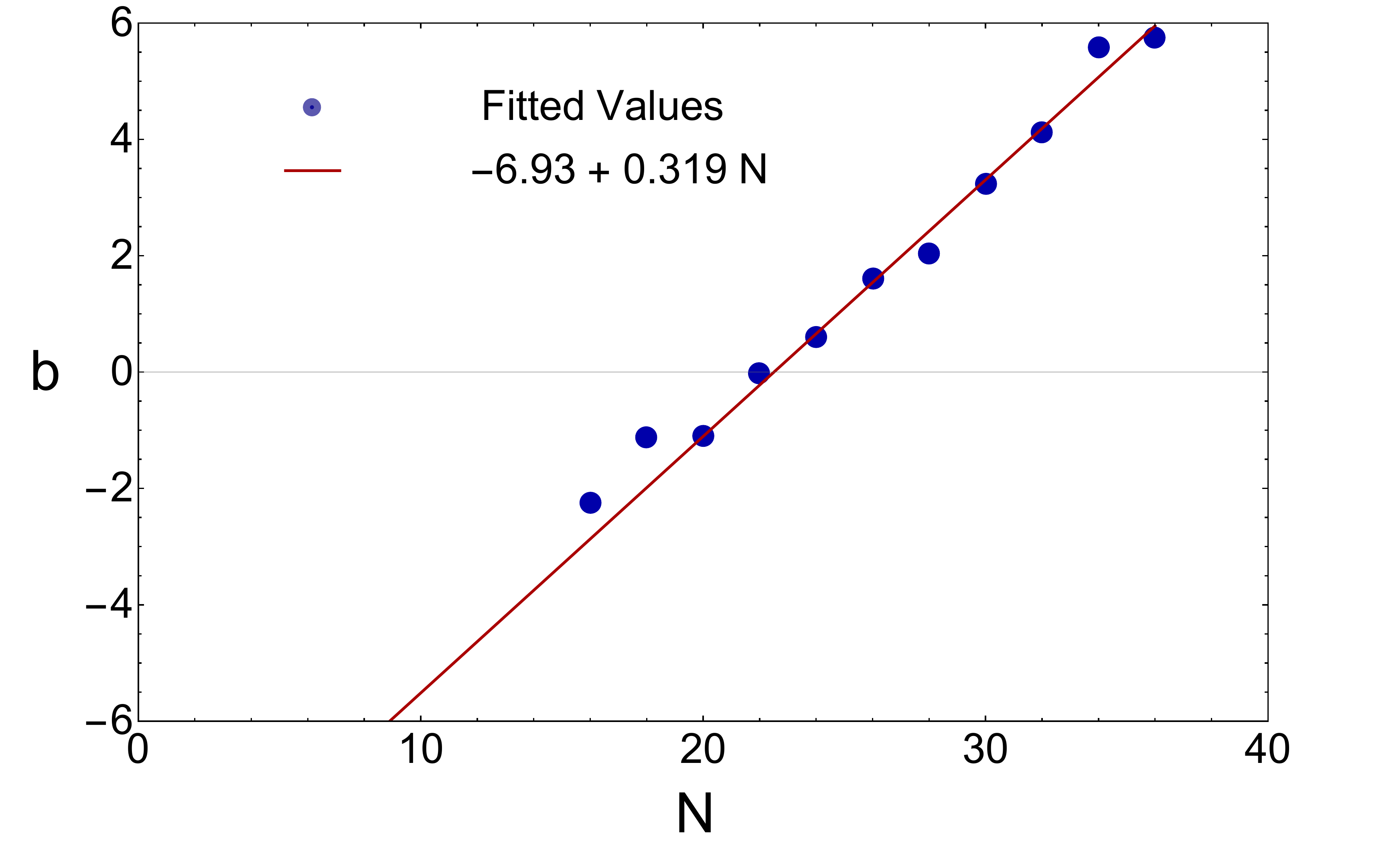}}
	\caption{The fitted values of $\log a$ (left) and $b$ (right), defined in Eq.(\ref{denfit}), versus $N$.
		The lines are the best fits to the data. In the right figure only
		the points for $N\ge 28$ have been used for the fitting}
	\label{fig:ab}
\end{figure}

In Fig. \ref{fig:ab} we depict  $\log a(N)$ and $b(N)$  by fitting of the exact partition function computed numerically by exact diagonalization.
The zero temperature entropy and the ground state energy are then obtained from Eq.(\ref{S0c}):
\be
S_0 = 0.21 N,  \qquad E_0 = -0.055 -0.029 N.
\ee
The value of $S_0$ is in rough agreement with the result $\sim 0.23 N$ obtained by 
Maldacena and Stanford \cite{maldacena2016}.

We now move to the calculation of the specific heat. 
In the very
low temperature limit with $\beta J \gg N$ we can expand the partition
function as
\be
Z(\beta) = \frac {c_0}{\beta^{3/2}} e^{-\beta E_0 + S_0}\left[1 + \frac c{2\beta J}\right].
\ee
It would be tempting to also make  the identification
\be
\frac c{2 } &=& \frac 32 N \frac {db}{dN}, \nonumber
\label{S0d}
\ee
but in the parameter range we are looking at it is not justified to
expand the exponential. Rather, we determine the specific heat coefficient $c$
by directly fitting the $\beta$-dependence of the specific heat, 
 \be
\frac {dU(T)}{dT},
 \ee
where the internal energy per particle, $U(T)$, is defined in the usual way,
 \be
 U(T) = -\frac 1N \frac {d\log Z}{d \beta}.
 \ee

 Setting $J = 1$ for convenience, and using the low temperature expansion of the partition function given in
 Eq. \eref{zmal},
 \be \label{partq}
 Z(\beta) = \frac 1{\beta^q} e^{-\beta E_0 +S_0 + \frac c{2\beta}},
   \ee
    we find that
   \be
   \frac{dU}{dT} = \frac qN + \frac cN T,
   \ee
where the exponent $q$ that controls the one-loop quantum correction $\beta^{q}$ to the partition function is left as a free parameter rather than fixing it to the perturbative \cite{kitaev2015,maldacena2016} prediction $ q= 3/2$. 

 In terms of the eigenvalues $E_{k,p}$ of the $p$'th
 member of the ensemble of SYK Hamiltonians, the specific heat per particle
 is given
 by
\be
\frac {dU(T)}{dT} = \frac 1{N} \frac 1Z \sum_{k,p} \frac{(E_{k,p} - \langle E\rangle)}{T^2}e^{-\beta E_{k,p}},
\label{c1}
  \ee
 with
 \be
 \langle E \rangle  = \frac 1{Z }\sum_{k,p} E_{k,p} e^{-\beta E_{k,p}},
 \ee
 and
 \be
Z  = \sum_{k,p}e^{-\beta E_{k,p}}.
 \ee
 For a given realization of the random Hamiltonian, the fluctuations of the average energy, 
 \be
 \bar E_p =\frac{ \sum_{k} E_{k,p} e^{-\beta E_{k,p}}}{ \sum_{k}  e^{-\beta E_{k,p}}},
 \ee
give rise to significant
finite size contributions to the specific heat  which can be eliminated by performing the ensemble average relative to the average energy for each
realization of the SYK Hamiltonian, i.e.,
\be
\frac {dU(T)}{dT} = \frac 1{N}\frac 1Z \sum_{k,p} \frac{(E_{k,p} - \bar E_p )^2}{T^2}e^{-\beta E_{k,p}}.
\label{c2}
 \ee
 For a large number of particles this procedure should be equivalent to the calculation according to Eq. \eref{c1}.
However, for the values of $N$ we work with, this  finite size effect must be removed in order to obtain accurate results for the low temperature limit of the
specific heat.

\begin{figure}[t!]
   	\centering
   	\includegraphics[width=7cm]{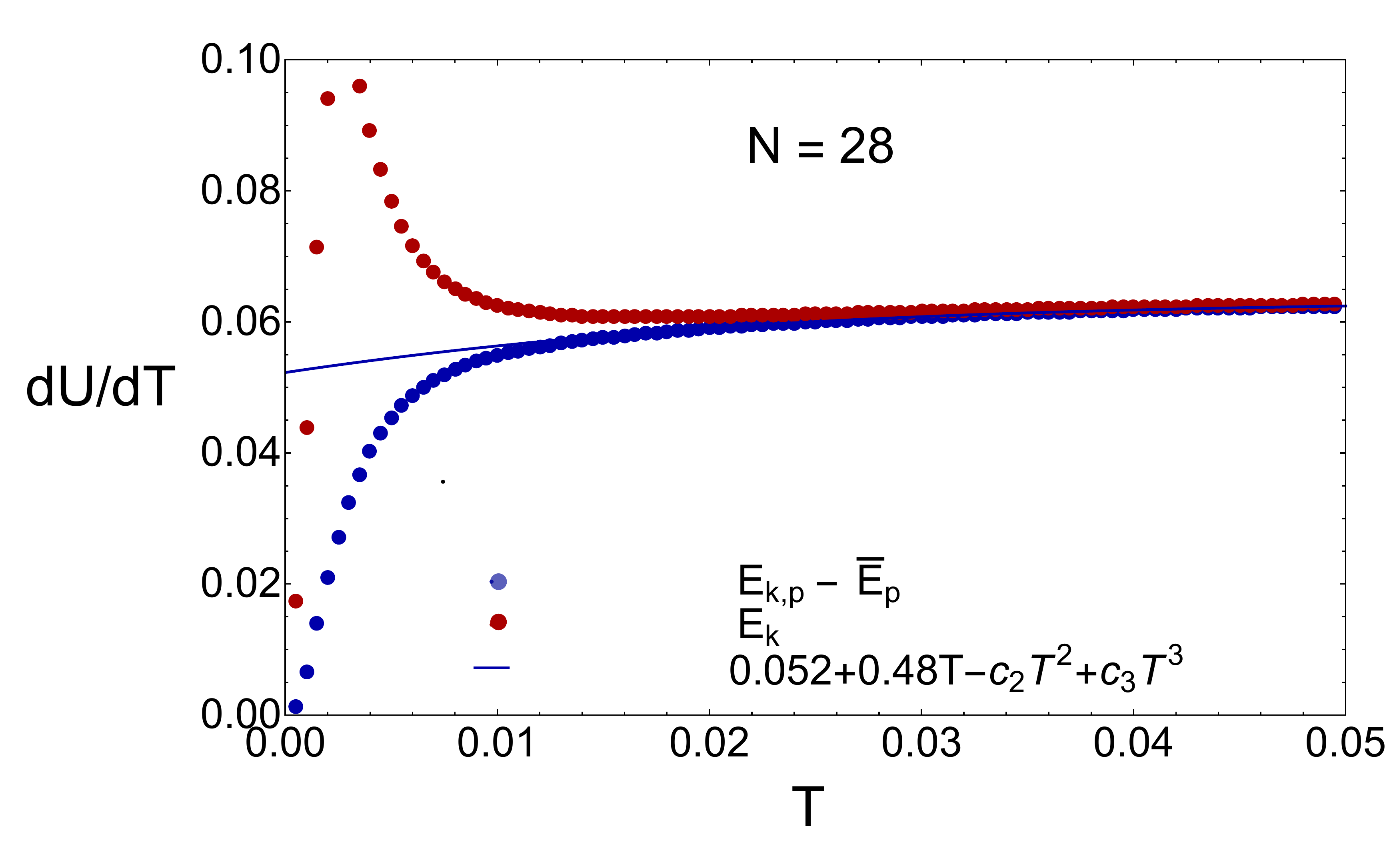}
   	\includegraphics[width=7cm]{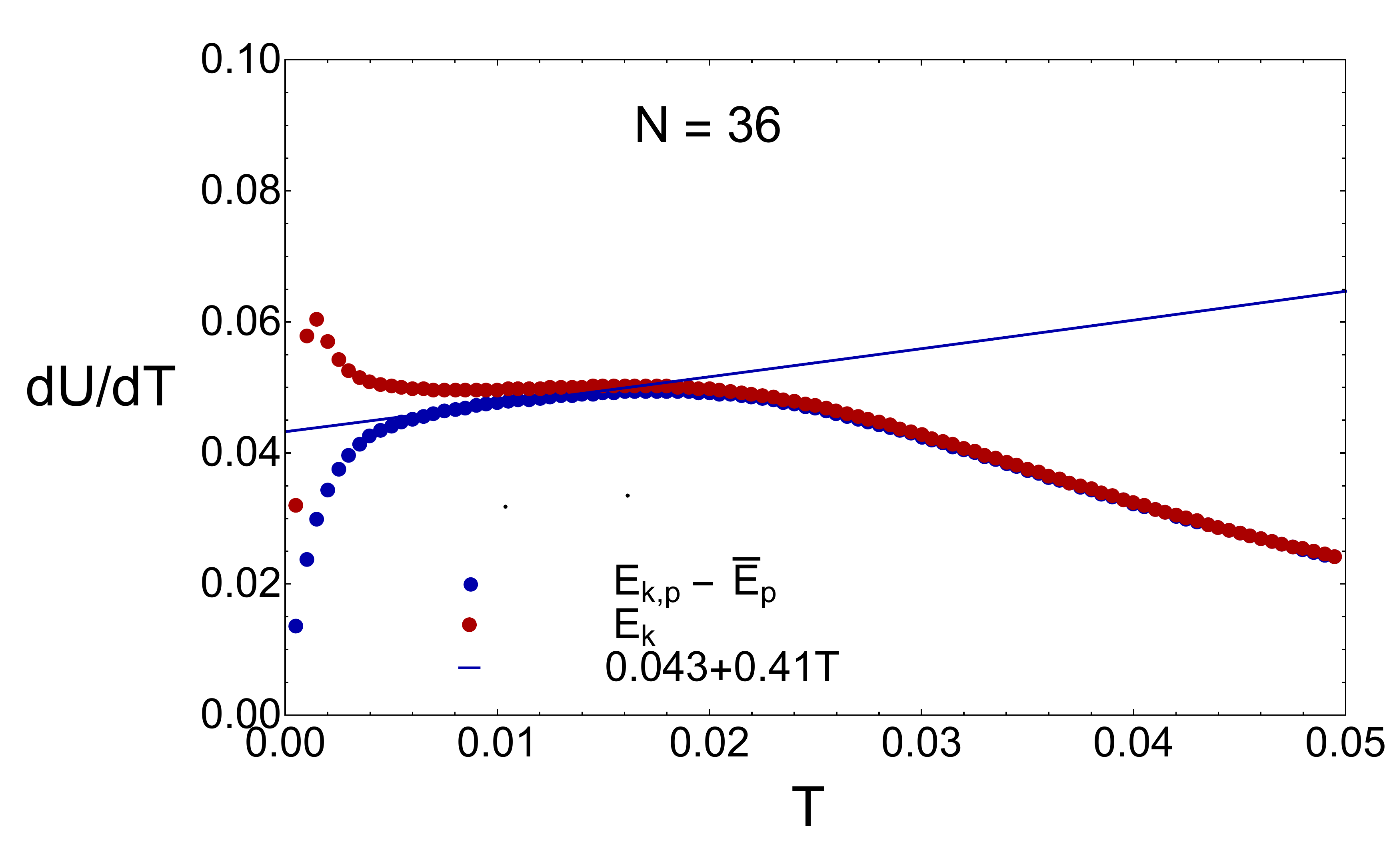}
   	\caption{The specific heat as a function of the temperature for $N=28$ (left) and $N=36$ (right).
   	  The red dots represent the numerical result for the SYK model when
          specific heat is calculated relative to ensemble average
          (see Eq. \eref{c1}) while the blue dots show the results where
          the free energy is calculated
          relative to the average energy $\bar E_p$ for each
          realization of the ensemble (see Eq. \eref{c2}).
 The blue curve is a linear fit to
   		the blue dots on the interval [0.0075, 0.015] for $N=36$ and cubic polynomial fit on [0.025,0.05] for $N=28$.}
   	\label{zbet}
   \end{figure}

 The finite size effects discussed in the previous paragraph decrease
 rapidly with the total number of particles. As an example we show
  in Fig.~\ref{zbet} the temperature dependence of the specific heat for $N = 28 $ (left) and
   $N=36$ (right). We show both the result where the specific heat
   is calculated according to Eq. \eref{c1}
   (red dots) and the result where we first calculate the specific
   heat for each realization of the Hamiltonian and then perform the ensemble
   average as given in Eq. \eref{c2} (blue dots).
   The curves are  fits to the blue dots. 

  Except for $N=36$, where we have only $2000$ eigenvalues
   for each configuration and  use a linear fit on a shorter fitting interval,  we use cubic fits
   \be \label{cn}
   \frac{dU(T)}{dT} = \frac {q(N)}N +c(N) T +c_2(N) T^2 + c_3(N) T^3. 
   \ee

     \begin{figure}[t!]
    	\centering
    	\includegraphics[width=7cm]{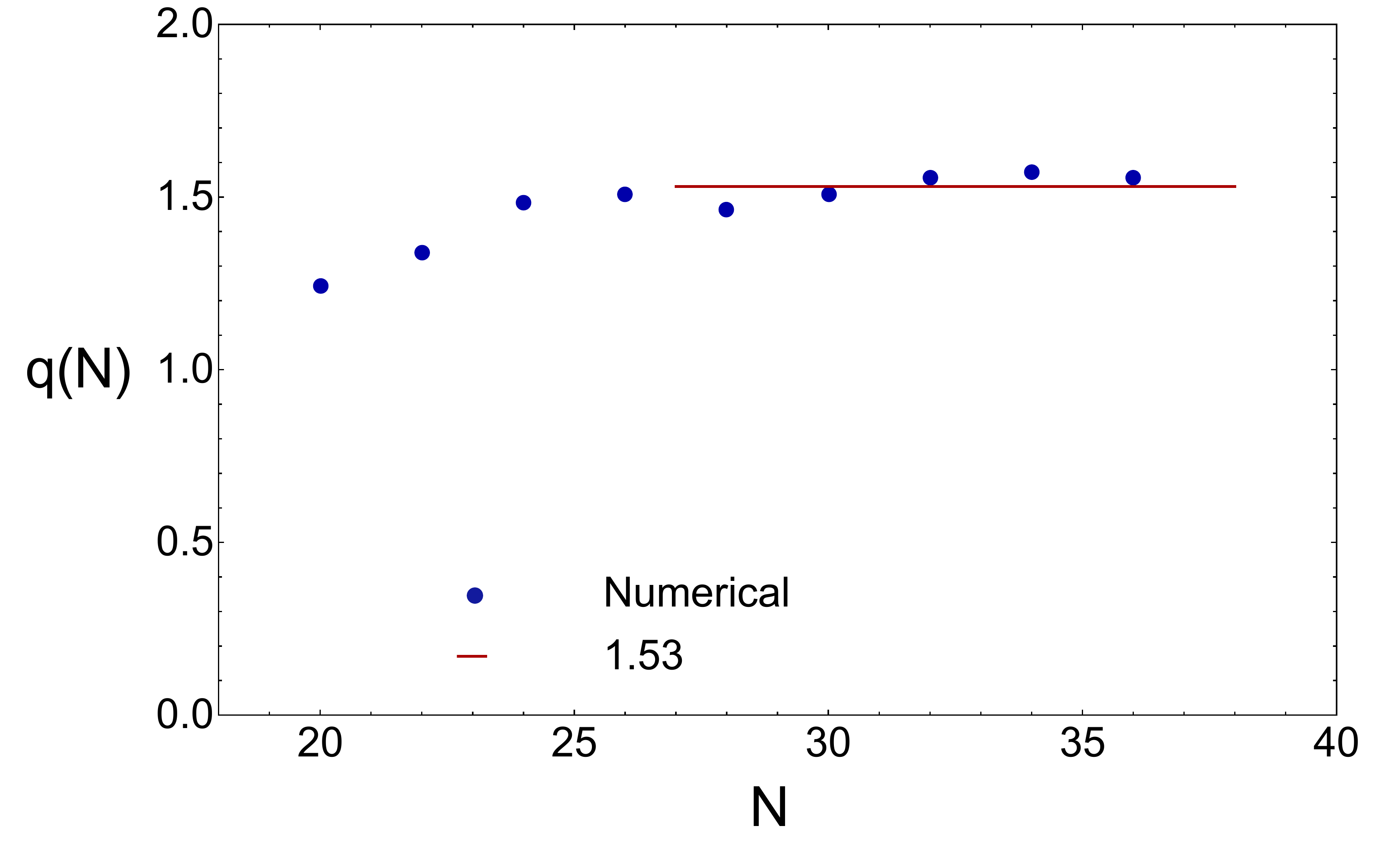}
    	\includegraphics[width=7cm]{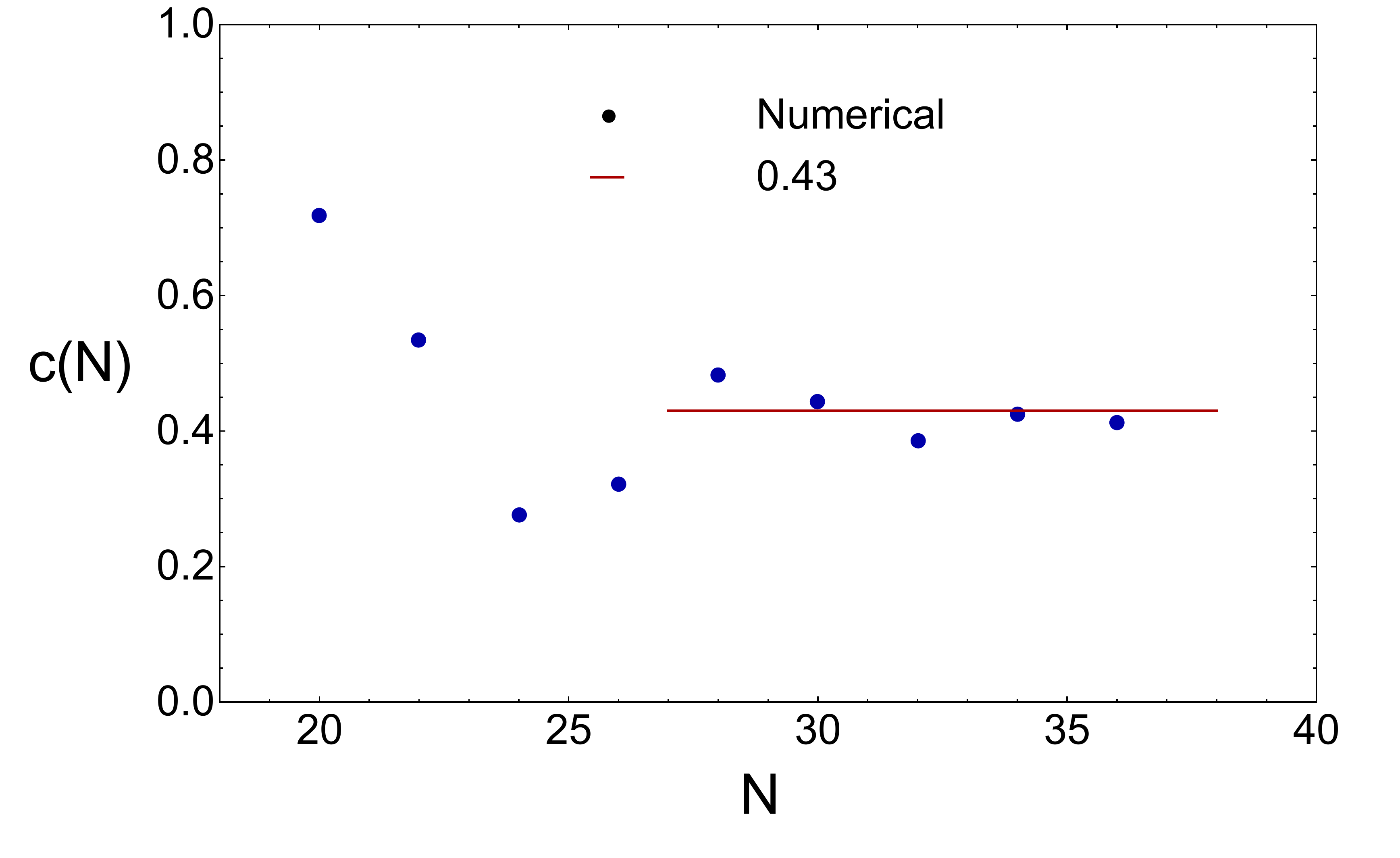}
    	\caption{The exponent of the prefactor, $\beta^{-q(N) } $ of the partition function (\ref{partq}) versus $N$ (left) and 
    		the specific heat coefficient $c(N)$, from (\ref{cn}), versus $N$ (right). The exponent $q(N)$ and the specific heat coefficient 
    	 $c(N)$ are fitted by a constant.}
    	\label{ab}
    \end{figure}

     In Fig. \ref{ab} we show the $N$-dependence of $q(N) $ (left) and $c(N)$ (right)
     which are fitted by a constant for $N \ge 28$ (see curves). This results in the following estimates
   for the exponent $q$ in Eq.\ref{partq} that controls one-loop quantum corrections and the specific heat coefficient
   \be
   q = 1.53 \pm 0.20, \qquad c/N = 0.43 \pm 0.10.
   \ee

   The value of $q$ is consistent with the estimate $q = 3/2$ \cite{maldacena2016} from an analytical calculation of one-loop quantum corrections to the classical action.
   It is also in agreement with the semicircular form of the spectral density, see
   Eq. \eref{denfit}. 
   Likewise the analytical estimation of the specific heat coefficient $c/N = 0.396$ \cite{maldacena2016} is also consistent with our numerical results.

We note that all the results of this section are based on the ansatz Eq. (\ref{denfit}) for the density of states. The exponent $1/2$ of the prefactor was chosen because it gave the best fit to the numerical results. However there is an indirect theoretical justification for that exponent.
In the recent literature on the SYK model there are several studies \cite{kitaev2015,maldacena2016,polchinski2016,jevicki2016} of the one-point temporal correlation function which is the Fourier transform
of the strength function
\be
-\frac 1N \sum_\alpha \sum_k |\langle 0|\gamma_\alpha|k\rangle|^2 \delta(E +E_k-E_0),
\ee
where $E_0$ is the $N$-particle ground state energy, $|k\rangle$ are eigenstates with $N\pm 1$
particles and $\gamma_\alpha$ is an Euclidean $\gamma$ matrix. These results are based on perturbative semi-classical techniques that typically are valid only up to time scales of the
order of the Ehrenfest time. However in \cite{bagrets2016} a non-perturbative treatment of quasi-zero modes enlarged the time domain of applicability of the analytical results to scales shorter but of the order the Heisenberg time. Interestingly, it was found \cite{bagrets2016} that, in an energy representation, the strength function for low energies $\propto \sqrt{E-E_0}$.
In principle the strength function is unrelated to the many-body spectral density Eq. (\ref{denfit}) because the former provides also information of the correlations between eigenvalues and eigenvectors. However,
if the eigenvectors and the eigenvalues are uncorrelated, as is the 
case for the
Wigner-Dyson random matrix ensembles, the strength function is proportional to the
spectral density.
Below we will see spectral correlations of the SYK model are well described by the Wigner-Dyson ensembles which justifies a posteriori the ansatz Eq. (\ref{denfit}) for the tail of the spectral density.

In summary, we have shown that the spectral density of the SYK model is Gaussian in
the  limit of a large number of particles $N$ so it is qualitative different from the semi-circle law typical of random matrices. However for a fixed finite $N$, the tail of the spectral density is close to a semi-circle law while the center is Gaussian.
The value of the zero temperature entropy and specific heat coefficient, obtained numerically from the tail of the spectrum and the low-temperature behavior of the partition function, are close
to previously obtained analytical estimates \cite{maldacena2016,jevicki2016}.

 \section{Spectral correlations}
 In this section we investigate eigenvalue correlations that provide valuable information on the dynamics of the system. 
 We focus on long time scales of the order of the Heisenberg time $\sim \hbar/\Delta$ where $\Delta$ is the mean level spacing.
 Disordered metals, or quantum chaotic systems, are expected to be described by
 the invariant random matrix ensembles in this region. 
 Physically, agreement with random matrix theory predictions indicates that an
 initially localized wave packet reaches the boundary of the sample for sufficiently long time scales.
 For a disordered insulator we expect level correlations to be described by Poisson statistics.
 Although in the literature on $k-$body embedded fermionic ensembles there are some reports of
 Poisson statistics for two-body random interactions in the dilute limit \cite{benet2001}, there is broad evidence from numerical and analytical findings \cite{srednicki2002,verbaarschot1984,bohigas1971a} that level statistics are very close to the random matrix theory prediction at least for short-range eigenvalue correlations.

 As was mentioned in the introduction, the only previous study of spectral correlations in the SYK 
 model \cite{you2016} investigated numerically the
ratio of consecutive level spacings
which only explores time scales of the order of the Heisenberg time.
For shorter time scales, corresponding to energy scales beyond
the mean level spacing, level statistics for the SYK model is yet an open problem. We shall see that level statistics in this region are well described by random matrix theory though deviations, that decrease with $N$, are systematically observed for larger spectral distances corresponding to time scales much shorter than the Heisenberg time. 

  The universality class for the spectral correlations is determined by the anti-unitary and involutive
  symmetries  of the system. Since the SYK Hamiltonian does not have any involutive symmetries, the
  universality class is given by the Wigner-Dyson random matrix ensembles with a Dyson index
  $\beta_D = 1$, 2 or 4. The first case is when the anti-unitary symmetry squares to one, the
  second case when there are no anti-unitary symmetries, and the third case when the anti-unitary
  symmetry squares to -1. The SYK Hamiltonian has two anti-unitary symmetries (See Table I)
  \be
     [C_1K, H_{SYK}] = 0, \qquad  [C_2K, H_{SYK}] = 0.
     \ee
     which is equivalent to one irreducible anti-unitary symmetry, $C_1 K$, and the unitary symmetry
     $C_1 K C_2 K$. Physically, the symmetries $C_1K$ and  $C_2K$ are charge conjugation symmetries which are equal
     to the product of the ``even'' gamma  matrices or ``odd'' gamma matrices, respectively (choosing
     the right labeling for ``even'' and ``odd''). Therefore,  $C_1 K C_2 K \sim \Gamma_5$ with
     $\Gamma_5={\rm diag}(1,\cdots,1,-1,\cdots,-1)$ in a chiral representation of the Dirac
     $\gamma$ matrices. In this representation the SYK Hamiltonian splits into two diagonal block matrices
     of equal size.
     If $C_1 K C_2 K=\pm\Gamma_5$, the charge conjugation matrix commutates with the projection on
     the diagonal blocks. If $(C_1K)^2 = 1$ it is possible \cite{porter} to find an $H$-independent basis for which the
     blocks become real, corresponding to a Dyson index $\beta_D = 1$. Moreover, if $(C_1K)^2 = -1$,
     it is possible to construct an $H$-independent basis for which the Hamiltonian can be arranged into
     quaternion real matrix elements corresponding to a Dyson index $\beta_D = 4$. 
     If $C_1 K C_2 K=\pm i \Gamma_5$,
     the charge conjugation matrix does not commute with the projection onto
     the blocks. Therefore we cannot use these symmetries to construct a basis for which the Hamiltonian becomes
     real or quaternion real. Since there are no unitary symmetries the matrix elements of the SYK Hamiltonian
     are complex corresponding to a Dyson index $\beta_D = 2$. However, the symmetry $C_1K$ still can be
     used to show that both blocks have the same eigenvalues (see \cite{kieburg2015} for a similar reasoning).
     We refer to Appendix A for all technical details.
      
     \begin{table}[t!]
   \begin{tabular}{c|c|c|c|c}
       $N$   &$ (C_1 K)^2$ & $(C_2K)^2$ &$C_1KC_2K$ &   RMT\\
       \hline
       2  &     1   &       -1    & $-i\Gamma_5$  & GUE \\
       4  &    -1   &       -1    & $-\Gamma_5 $   & GSE \\
       6  &    -1   &        1    & $-i\Gamma_5$   & GUE \\
       8  &     1   &        1    & $\Gamma_5  $   & GOE \\
       10 &     1   &       -1    & $-i\Gamma_5$   & GUE \\
       12 &    -1   &       -1    & $\Gamma_5$     & GSE \\
        \end{tabular}
   \caption{(Anti-)Unitary symmetries of the SYK Hamiltonian and the corresponding random matrix ensemble.
     The symmetries are periodic in $N$ modulo 8.}
     \end{table}
 For our study we employ the level spacing distribution $P(s)$ (\ref{eq:spacing_distribution}), the probability to find two neighboring eigenvalues separated by a distance $s=(E_{i+1}-E_i)/\Delta$, and the number variance $\Sigma^2(L)$ (\ref{eq:nv}), that describes fluctuations in the number of eigenvalues in a spectral window of size $L$ again measured in units of the mean level spacing $\Delta$. The latter, a long-range spectral correlator directly related to the two-point correlation function, gives information on the quantum dynamics for times scales of the order but much larger than the mean level spacing (Heisenberg time). We shall use it to investigate deviations from random matrix predictions. The former is more suited to study longer time scales $\approx \hbar /\Delta$ and also provides indirect information on higher order correlation functions. 
 
 We investigate level statistics numerically by an exact diagonalization of the upper block of
 the Hamiltonian (\ref{hami}) for $N \leq 36$. 
 The first step in the spectral analysis is the unfolding of the spectrum \cite{guhr1998}, namely, to rescale the spectrum so that the mean level spacing is the same for all energies. This is a necessary condition to compare level statistics in different parts of the spectrum. For that purpose, for each $N$, we employ the averaged smooth staircase function (the integral of the spectral density) 
 resulting from a fifth order polynomial fitting involving only odd powers, to unfold the spectrum.
 The spectrum rescaled in that way, which has unit mean level spacing for all energies, is ready for the level statistics analysis. We have observed that level statistics are similar for all energies. Except for $N = 36$, where we have only obtained about $2\%$ of eigenvalues close to the edge of the spectrum, we have taken about $70\%$ of the eigenvalues around $E \approx 0$.  
\subsection{Short range spectral correlations: $P(s)$} 
The level spacing distribution $P(s)$ is the probability to find two eigenvalues separated at a distance $s$ in units of $\Delta$ with no other eigenvalues in between:
\begin{equation}
P(s) = \sum_i \left\langle \delta(s-\epsilon_i+\epsilon_{i+1})\right\rangle \ \ \  \epsilon_i = E_i/\Delta.
\label{eq:spacing_distribution}
\end{equation}
\begin{figure}%
	\centering
	\resizebox{0.7\textwidth}{!}{\includegraphics{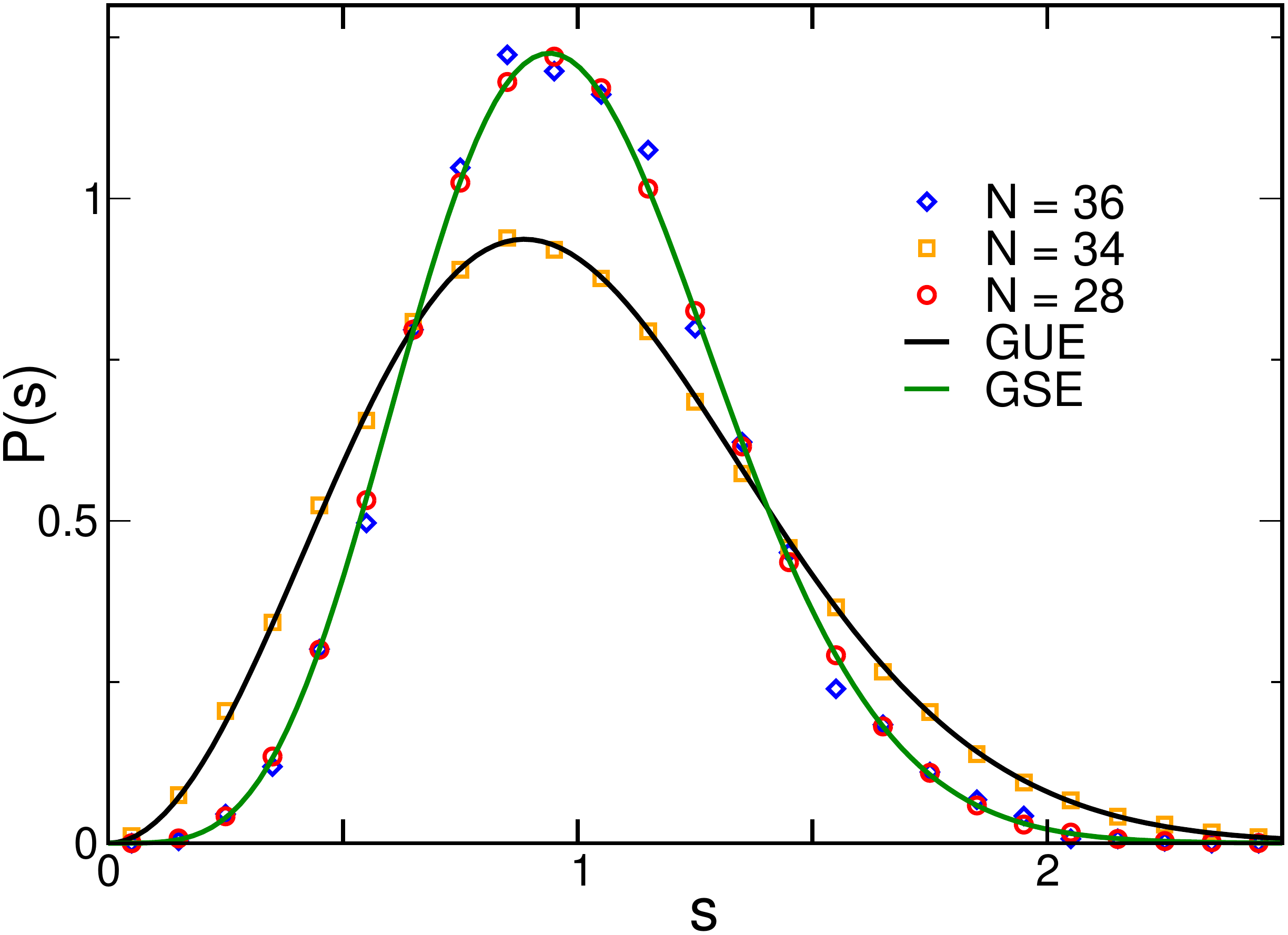}}
	\caption{Level spacing distribution $P(s)$ Eq. (\ref{eq:spacing_distribution}). Numerical results, (circles) $N = 28$, (squares) $N = 34$ and (diamonds) $N = 36$, are in excellent agreement with the predictions of random matrix theory (solid lines) corresponding to the Gaussian Unitary Ensemble (GUE) for $N = 34$ and the Gaussian Symplectic Ensemble (GSE) for $N = 28,\; 36$. We note that while for $N = 28,\; 34$ we have taken about $70\%$ of the available spectrum around the center of the band, for $N = 36$, where we cannot diagonalize the full Hamiltonian, we consider only a total of about $15000$ different eigenvalues close to the ground state from $15$ disorder realizations. The universality class is controlled by the type of allowed representations of the Clifford algebra of the Majorana fermions which is sensitive to $N$ (see Table I and the main text for more details). These results clearly show that the SYK model has quantum chaotic features even for large times $s \sim 1$ of the order of the Heisenberg time.}
	\label{fig4a}
\end{figure}
\begin{figure}[b]%
	\centering
	\resizebox{0.7\textwidth}{!}{\includegraphics{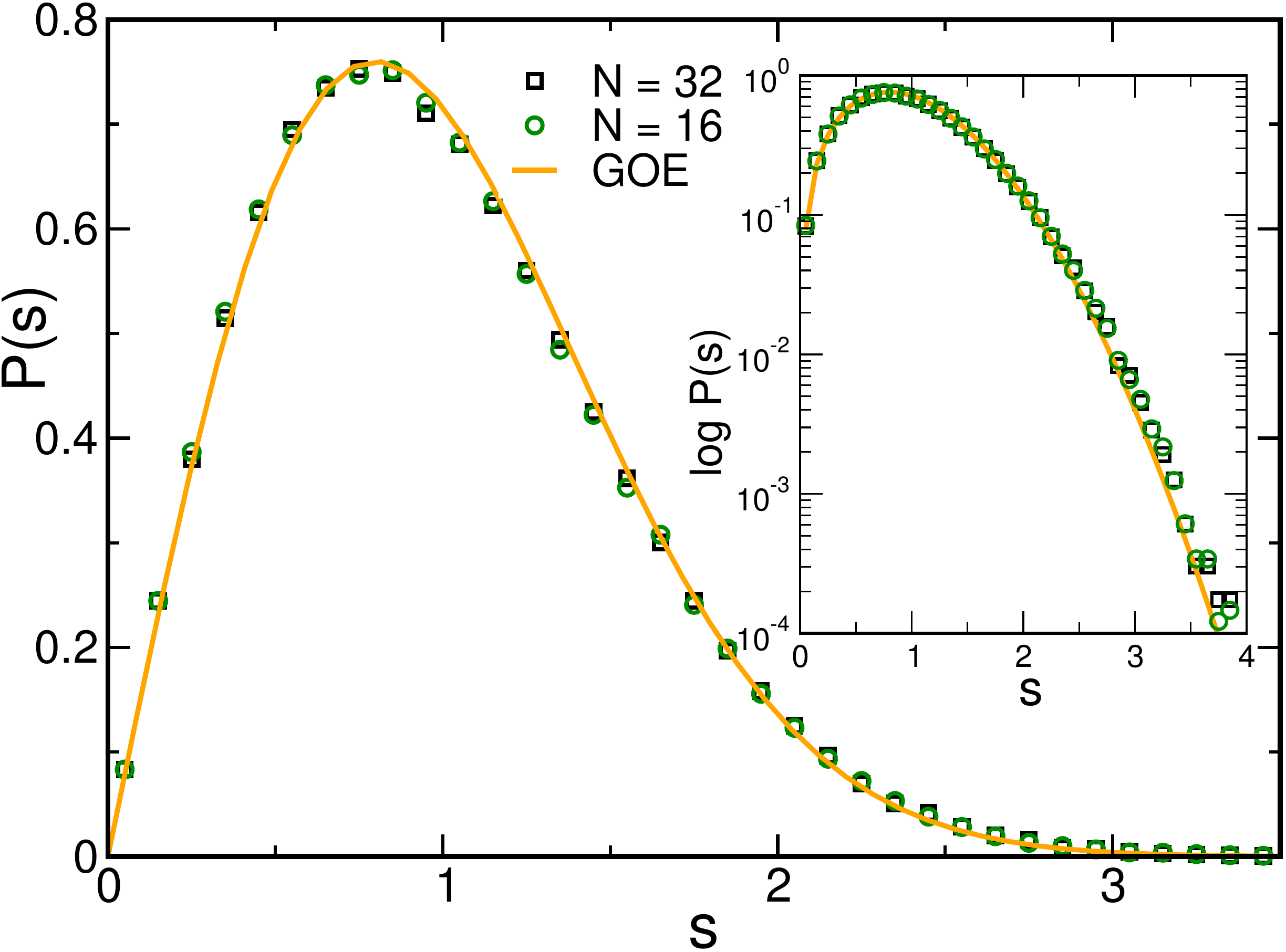}}
	\caption{Level spacing distribution $P(s)$, Eq. (\ref{eq:spacing_distribution}), for $N = 16$ (circles)  and $N = 32$ (squares). For both dimensions the Clifford algebra admits a real representation so the expected universality class is that of the Gaussian Orthogonal Ensemble (GOE). Indeed  the numerical results (symbols) are in excellent agreement with the GOE prediction. Interestingly, despite the vast difference in size, we do not observe substantial deviations from the GOE prediction even for $N = 16$ where mesoscopic fluctuations are expected to be stronger.}
	\label{fig5}
\end{figure}
In an insulator it is given by Poisson statistics:
$
P(s) = e^{-s}.$
By contrast, the random matrix prediction, that applies to a disordered metal and to a quantum chaotic system, is very well approximated by the Wigner surmise, 
\begin{equation}
P(s) \approx a_\beta s^\beta\exp(-b_\beta s^2).
\label{eq:wd}
\end{equation}
Level repulsion, $P(s) \to 0$ for $s \to 0$, is a distinguishing feature of extended states though its strength depends on the global symmetries of the Hamiltonian (\ref{hami}). For systems that admit a real representation of the Hamiltonian,
due to time reversal invariance (or more generally due to an anti-unitary symmetry that
squares to 1), $\beta =1,\; a_1 = {\pi}/{2},\; b_1= {\pi}/{4}$. Similarly if the Hamiltonian only admits a complex representation, due for instance to the breaking of time translational invariance as a consequence of a magnetic field or flux,
$\beta =2,\; a_2 = {32}/{\pi^2},\; b_2= {4}/{\pi}$.
Finally the case  $\beta = 4,\; a_4 = {262144}/{729\pi^3},\; b_4= {64}/{9\pi}$
corresponds to systems with time-reversal symmetry and strong spin-orbit interactions leading to a doubly degenerate spectrum 
(or more generally to systems with an anti-unitary symmetry that squares to $-1$).
It is typical of random matrices with quaternionic entries.

In Fig.~\ref{fig4a} we plot $P(s)$ for $N = 28,\;N = 34$ and  $N = 36$.
Excellent agreement with the random matrix prediction is found in all cases. As can be seen from Table I,
$N=28$ belongs to the Gaussian Symplectic Ensemble (GSE) universality class ($\beta_D = 4$), while
$N = 34$ belongs to the Gaussian Unitary Ensemble (GUE) universality class ($\beta_D = 2$).
We note that the $N$ dependence of the universality class was already reported in \cite{you2016}, although it was not  discussed that this was a simple consequence of two features of Clifford algebras: the existence of real, complex or quaternionic representations for different values of the dimensionality $N$ and Bott periodicity, namely, these representations follow a periodic pattern, in this case the Bott periodicity is $N\, {\rm mod} 8$. An example of a period is: $N = 36$: GSE, $N = 34$: GUE, $N = 32$: GOE, $N = 30$: GUE, and so on.

In Fig.~\ref{fig5} we depict $P(s)$ for $N = 16$ and $N= 32$ both belonging to the Gaussian Orthogonal Ensemble (GOE) universality class. 
Even though the large difference in size, we do not observe important differences between the two cases. 
We will see in the following analysis of the number variance, a long-range spectral correlator that deviations from random matrix theory eventually occur for larger eigenvalue separations which indicates that the SYK model is not ergodic for sufficiently short time scales.
\subsection{Long range spectral correlations: the number variance $\Sigma^2(L)$} 
The number variance is defined as the variance of the number of levels $N$ inside an energy interval that has (in units of the mean level spacing) $L$ eigenvalues on average:

\begin{equation}
\Sigma^2(L) = \left\langle N^2(L)\right\rangle - \left\langle N(L)\right\rangle^2.
\label{eq:nv}
\end{equation}

For a Poisson distribution typical of an insulator, different parts of the spectrum are not correlated, so the number variance is linear with slope one, $\Sigma^2(L) = L$. 
\begin{figure}%
	\centering
	\resizebox{0.85\textwidth}{!}{\includegraphics{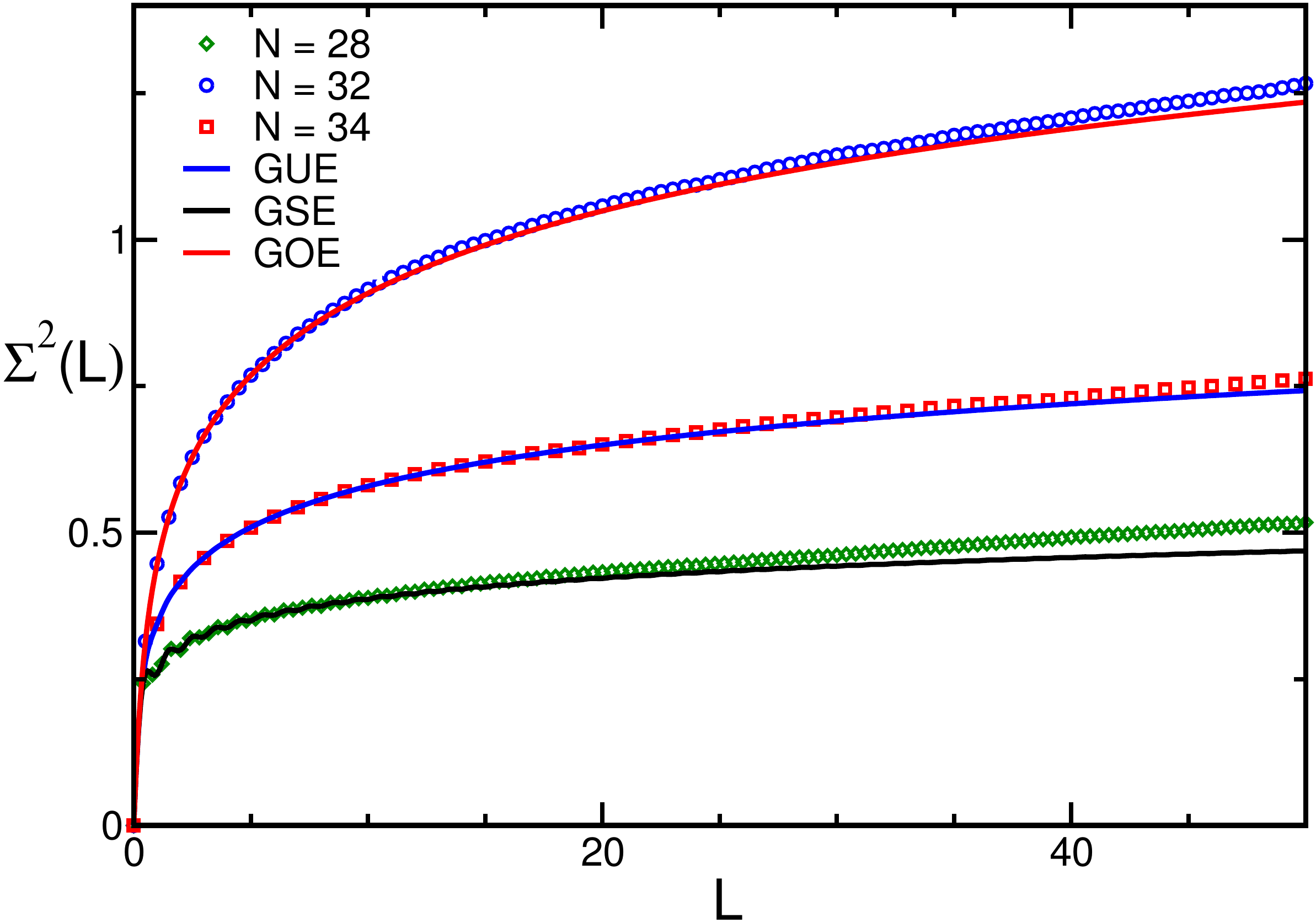}}
	\caption{Number variance $\Sigma^2(L)$, Eq. (\ref{eq:nv}), for $N = 28$ (diamonds), $N = 32$ (circles), $N = 34$ (squares), which belongs to the GSE, GOE and GUE universality classes respectively (solid lines), as a function of the width $L$ of the energy interval, corresponding to a spectral window with $L$ eigenvalues on average.
		For sufficiently large $L$,  we start observing deviations from the random matrix theory predictions.}
	\label{fig6}
\end{figure}
The random matrix prediction, that also occurs in non-interacting \cite{altshuler1988,braun1995} and strongly coupled \cite{bertrand2016} disordered metals below the Thouless energy, is that level repulsion causes, for $L \gg 1$, a slow logarithmic increase, usually termed {\it level or spectral rigidity} of the number variance:
\begin{equation}
\Sigma^2(L) \approx c_\beta(\log(d_\beta \pi L)  + \gamma +1 +e_\beta \ldots),
\label{eq:nvwd}
\end{equation}
\begin{figure}[t]%
	\centering
	\resizebox{0.85\textwidth}{!}{\includegraphics{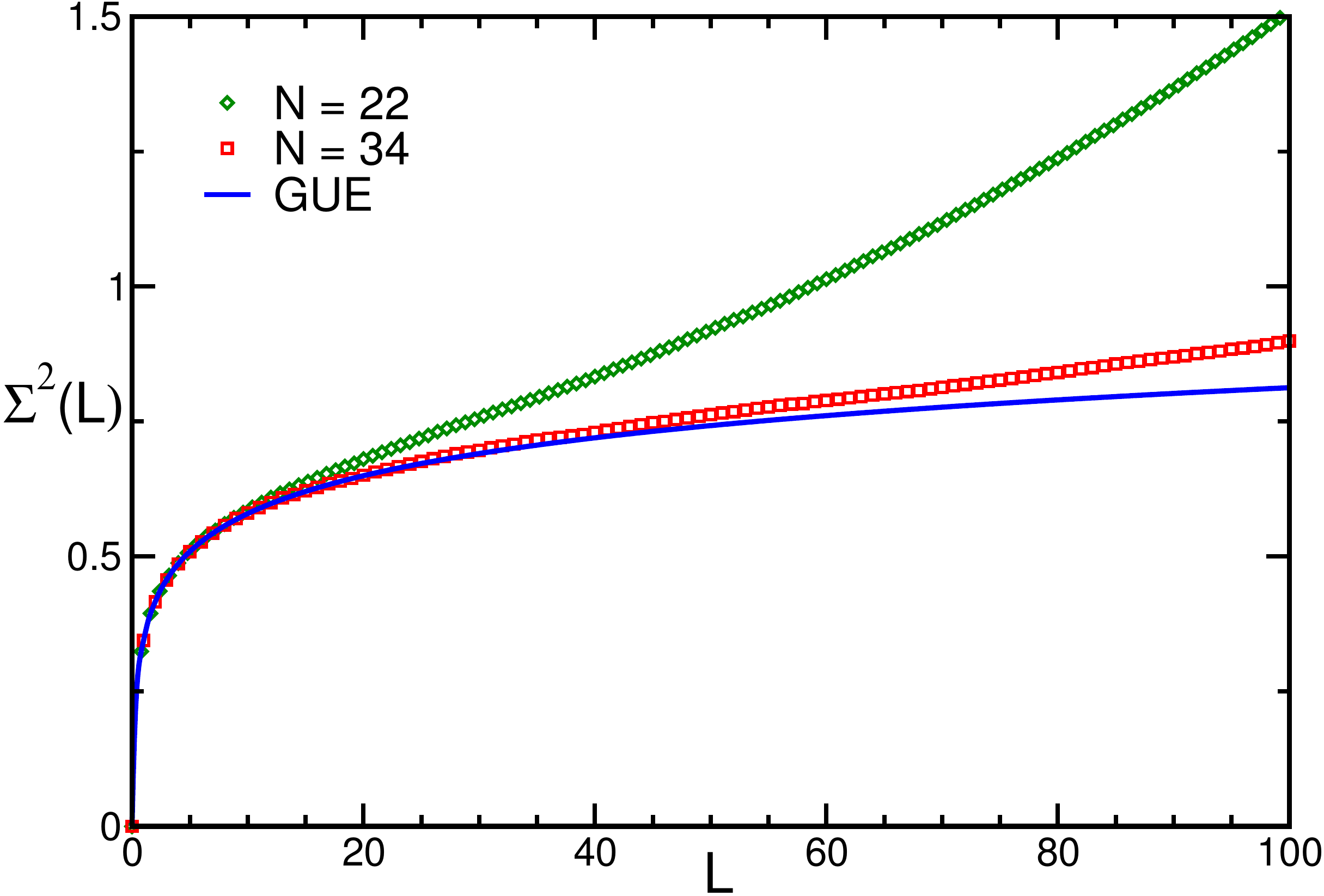}}
	\caption{Number variance $\Sigma^2(L)$, Eq. (\ref{eq:nv}),
          for $N= 22$ and $N =34$, both corresponding to the GUE universality classes. We observe that deviations from the random matrix prediction occur much earlier for the smaller dimension $N=22$. This suggests that the observed deviations are due to mesoscopic fluctuations in a way reminiscent to the existence of a Thouless energy in the system. It is also an indication that the system is not ergodic and chaotic for sufficiently short times, an expected feature \cite{maldacena2016a,kitaev2015} in field theories with a gravity dual.}
	\label{fig7}
\end{figure}
with $c_1={2}/{\pi^2},\;c_2=c_1/2,\; c_4 = c_1/4$,\; $d_1 = d_2 = 2,\; d_4 = 4$,\; $e_1 = -\pi^2/8,\; e_2 = 0,\; e_4 = \pi^2/8$ and $\gamma = 0.5772 \ldots$ is Euler's constant. 
In Fig.~\ref{fig6} we depict the number variance for several values of the system size, $N = 28$, $N=32$ and $N=34$, each of them belonging to a different universality class: GOE for $N = 32$, GUE for $N = 34$ and GSE for $N = 28$. For all universality classes we find an excellent agreement with the random matrix prediction for small $L$. However we observe systematic deviations for sufficiently large $L \gtrsim 30$. As $N$ increases the region of agreement with random matrix increases as well, namely, deviations are observed only for larger $L$.

In Fig.~\ref{fig7} we depict the number variance for two sizes ($N = 22$ and
$N=34$) belonging to the same universality but one matrix size much smaller than the other. The idea is to study finite size effects, related to mesoscopic fluctuations in the number variance.
For small $L \leq 20$ the number variance follows the GUE prediction for both sizes. However for larger $L$, deviations from the random matrix result occur much earlier, and grow much faster, for $N = 22$  than for $N = 34$. 
An eyeball estimate suggests that the region of agreement with random matrix predictions scales approximately as $2^{N/8}$. 

Several conclusions  can be drawn from these results: 
a) the SYK model has spectral correlations similar to that of a disordered metal or a quantum chaotic systems even for energy scales much larger than the inverse mean level spacing, b) deviations for sufficiently large scales, suggest that, unlike a dense random matrix, the SYK model is not ergodic for sufficiently short time scales. This is expected as the Hamiltonian is rather sparse with only $\sim N^4$ non zero elements. This feature is also required for a gravity-dual interpretation where it is expected that, for times of the order of the Ehrenfest time  $\sim \log 1/\hbar$, certain correlation functions grow exponentially at a rate controlled by the Lyapunov exponent of the system \cite{maldacena2015}, c) the fact that, as $N$ increases, deviations from random matrix occur for larger $L$ is a strong indication that the observed chaotic features persist in the thermodynamic limit. It also suggests the existence of the equivalent of a Thouless energy in the system related to the typical time necessary to explore the full available phase space. 

\section{Outlook and conclusions}

We have shown analytically that, in the limit of large number of particles,
the SYK Hamiltonian has a Gaussian spectral density, although for a fixed finite number of particles, we have found numerically the tail of the density is well approximated by the semicircle law. Level statistics are well
described by random matrix theory up to energy scales much larger, but still of the order, of
the mean level spacing. Deviations from random matrix theory for larger energies, or shorter
times, are an indication that the model is not ergodic for short times. Together with previous results,
this a further confirmation that the SYK model has quantum chaotic features at any time scale. According to \cite{maldacena2016}, this is an expected feature in field theories with a gravity-dual. Indeed, we have numerically calculated the specific heat and the entropy and
found that the low temperature thermodynamic properties of the
SYK model are similar to those of a gravity background with a AdS$_2$ infrared limit. 
To some extent, our work on the SYK model shows that
a compound nucleus may have a gravity dual.
Finally we mention a few venues for further research. It would be interesting to
explore metal-insulator transitions in the model by reducing the range of the interaction
from infinity to a power-law decay. Another interesting problem is to evaluate analytically
the two level correlation function in the $N \to \infty$ limit by the replica trick by following the procedure of \cite{verbaarschot1984} for the $k$-body embedded ensemble. Similarly, the analytical evaluation
of the leading finite $N$ corrections of the spectral density, by a careful evaluation of higher order $N$ moments, would provide a full description of the low temperature thermodynamic properties of the model. This is necessary step for a full understanding of the relevance of the SYK model in holography. 
We plan to address some of these problems in future publications.

\acknowledgments{This work  acknowledges partial support from
  EPSRC, grant No. EP/I004637/1 (A.M.G.) and U.S. DOE Grant
  No. DE- FG-88FR40388 (J.V.). We also thank the Galileo Galilei Institute for Theoretical Physics and the INFN (A.M.G.) as well as
  the program Mathematics and Physics at
  the Crossroads at the INFN Frascati (J.V.) for hospitality and
  partial support during
  the initial stages and the completion of this work.
 Aurelio Bermudez, Bruno Loureiro, Masaki Tezuka (A.M.G.) and Mario Kieburg (J.V.) are thanked for illuminating discussions.  A. M. G. also thanks Masaki Tezuka for sending
 slides about his forthcoming work on the spectral form
 factor of the SYK model. \\
 
    {\it Note added in proof.—}  After this paper was accepted for
    publication, a paper \cite{cotler2016} appeared that also studies thermodynamic
    and spectral properties in the SYK model. In this
    paper spectral correlations are investigated by means
    of correlators of partition functions which at infinite
    temperature reduce to the spectral form factor which is
    the Fourier transform of the two-point correlation function.
}
 
\appendix
\section{Construction of the $\gamma$ matrices}

The $\gamma$ matrices are constructed iteratively starting from the $\gamma$ matrices in two dimensions
\be
\gamma^{(2)}_1 = \sigma_1,\qquad \gamma^{(2)}_2 =\sigma_2,\qquad \gamma^{(2)}_3 =\sigma_3.
\ee
and using the recursion relation
\be
\gamma^{(d+2)}_k &=& \sigma_1 \otimes \gamma^d_k , \qquad {\rm for} \quad k=1,\cdots, d+1,\nn\\
\gamma^{(d+2)}_{d+2}& =& \sigma_2 \otimes 1_{2^{d/2}},
\ee
to extend it to $d+2 = N$ 
dimensions where $N$ is the even number of Majorana fermions.
As we will see below, in this representation, the product of four gamma matrices is block diagonal. 

We can construct two anti-unitary symmetry operators
(Note that the gamma matrices in $C_1$ are purely imaginary while the $\gamma$ matrices in $C_2$
are purely real.)
\be
C_1& = &\gamma_1 \prod_{i=2}^{N}\gamma_{2i} K,\nn\\
C_2&= & \gamma_2 \prod_{i=2}^{N-1} \gamma_{2i+1}K,\nn\\
\ee
where $K$ is the complex conjugation operator 
(we could have interchanged the labels of
$\gamma_1$ and $\gamma_2$ so that $C_1$ would have been the product of the odd gamma matrices and
$C_2$ the product of the even gamma matrices). 
They satisfy the symmetry relations
\be
C_1K \gamma_\mu + (-1)^{N/2} \gamma_\mu C_1K = 0, \qquad C_2 K\gamma_\mu - (-1)^{N/2} \gamma_\mu C_2K = 0
\ee
with $\mu = 1, \ldots N$. 
Since the Hamiltonian is a sum of  products of four $\gamma$ matrices, we have
\be
   [C_1K, H] = 0, \qquad [C_2K, H] =0.
   \label{commutes}
   \ee
   We also have that
   \be
      [C_1K,C_2K] =0.
      \ee
  
      In the above table, which was also given in the main text,
      we give the main properties of these anti-unitary symmetries.
     \begin{table}
   \begin{tabular}{c|c|c|c|c}
       $N$   &$ (C_1 K)^2$ & $(C_2K)^2$ &$C_1KC_2K$ &   RMT\\
       \hline
       2  &     1   &       -1    & $-i\Gamma_5$  & GUE \\
       4  &    -1   &       -1    & $-\Gamma_5 $   & GSE \\
       6  &    -1   &        1    & $-i\Gamma_5$   & GUE \\
       8  &     1   &        1    & $\Gamma_5  $   & GOE \\
       10 &     1   &       -1    & $-i\Gamma_5$   & GUE \\
       12 &    -1   &       -1    & $\Gamma_5$     & GSE \\
   \end{tabular}
   \caption{Anti-Unitary symmetries of the SYK model and the corresponding random matrix
     theory. This table is also given in the main text.}
       \end{table}
     Because of \eref{commutes} we have that $[\Gamma_5, H] = 0$, with $\Gamma_5 = i^{-N/2} \prod_{i=1}^{N} \gamma_i$, so that $H$ splits
     into two block-diagonal matrices of the same size. 
     If $C_1KC_2K=\pm \Gamma_5$, then
     \be
     P \equiv \frac 12(1+ C_1K C_2K)
     \ee
     is a projection operator
     \be
     P^2 = P,
     \ee
     and
     \be
        [P,H] = 0, \qquad [C_1K,P]=0.
        \ee
        In this case we have that $(C_1K)^2=(C_2K)^2=\pm 1$. If $(C_1K)^2 =1 $ it is possible to find
        an $H$-independent basis in which $H$ becomes real, and the corresponding random matrix ensemble
        is the Gaussian Orthogonal Ensemble (GOE). If $(C_1K)^2 =-1 $ the Hamiltonian is self-dual
        quaternion up to an $H$ independent unitary transformation which corresponds to the
        Gaussian Symplectic Ensemble. In this case the eigenvalues of $H$ are a multiple of the
        quaternion identity and are thus doubly degenerate.
        
        If $C_1KC_2K= \pm i\Gamma_5$ the projection operator is given
        by
        \be
        P_i = \frac 12( 1 \pm i C_1K C_2K),
        \ee
        and
        \be
           [P_i,H] =0,
           \ee
           but
        because of the ``$i$'' this projection operator does not commute with $C_1 K$ or
        $C_2K$. So there are no anti-unitary symmetries when $H$ is block-diagonal, and we are in the universality class of the Gaussian Unitary
        Ensemble. In this case the charge conjugation matrices anti-commute with $\gamma_5$,
        \be
        \{C_1,\Gamma_5\} =0, \qquad  \{C_2,\Gamma_5\} =0,
        \ee
        so that $C_1$ and $C_2$ are block off-diagonal
        \be
        C_1K = \bmat 0 & c_1 K \\ c_1^* K& 0 \emat, \qquad
        C_2K = \bmat 0 & c_2 K \\  c_2^*K & 0\emat
        \ee
        with $c_{1,2}^* c_{1,2}= - 1$.
        If
        \be
        H =\bmat A & 0 \\ 0 & B \emat
        \ee
        then the anti-unitary symmetries \eref{commutes} result in the relation
        \be
        B^* = -c_i^* A c_i, \qquad i = 1,2.
        \ee
        Because $A$ and $B$ are Hermitian and $c_i^* c_i = -1$ we find from the secular
        equation that $A$ and $B$ have the same eigenvalues.
        
\section{Calculation of the fourth and sixth Cumulant}

In this appendix we calculate the normalized fourth and sixth cumulant
for the Hamiltonian of the SYK model.

\subsection{The fourth cumulant}

\begin{figure}
            \centering
            \includegraphics[width=10cm]{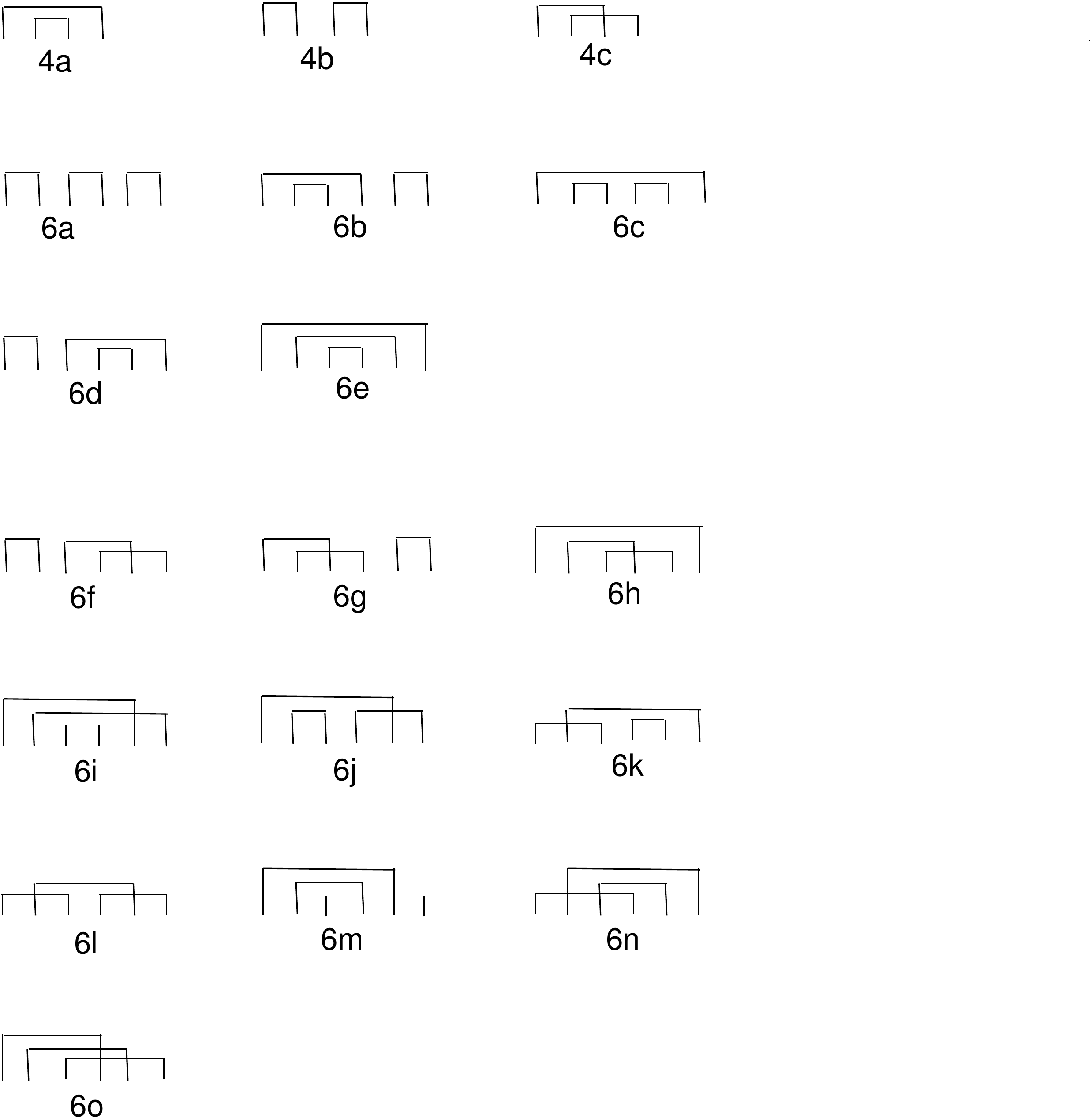}
            \caption{Contractions contributing to the 4th and 6th order cumulants.}
\label{fig:dia}
          \end{figure}

The normalized fourth cumulant is given by
\be
\kappa_4 = \frac {M_4(N)}{M_2^2(N)} - 3.
  \ee
  We now to proceed to the calculation of $M_4(N)$.
  The Gaussian average is the sum over all pairwise contractions. Because
  $\Gamma_\alpha^2 =1$ with $\Gamma_\alpha $ a product of four different gamma
  matrices we find that the nested contractions are given by
  \be
  2 M_2^2(N)
  \ee
  with the factor 2 corresponding to the two contractions 4a) and 4b) in Fig. \ref{fig:dia}.
  For the intersecting contraction, see Fig. \eref{fig:dia} (4c), we have to evaluate the trace
  \be
  \Tr\sum_{\alpha \beta} \Gamma_\alpha \Gamma_\beta \Gamma_\alpha \Gamma_\beta.
  \ee
  We have that
  \be
  \Gamma_\alpha \Gamma_\beta = (-1)^q  \Gamma_\beta \Gamma_\alpha
  \ee
  with $q$ the number of gamma matrices that $\alpha$ and $\beta$
  have in common.  For the sum over $\alpha$ and $\beta$ we thus
  obtain (see diagram 4c) in Fig. \ref{fig:dia}
  \be
  T_{4c}=  M_2^2(N){N\choose 4}^{-1}
    \sum_{q=0}^4(-1)^q
{N-4 \choose 4-q}{4\choose q}.
        \ee
        Note that, as a check of this result, that  without the
        factor $(-1)^q$ the sum over $q$ just gives
${N\choose 4}$.
          The result $T_{4c}$ can be simplified to
          \be
          T_{4c} = M_2^2(N)\frac{N^4 -38N^3+491N^2-2566N+4608}{N(N-1)(N-2)(N-3)}
          \ee
          This results in the normalized fourth order cumulant
           \be
          \kappa_4(N)&=&  \frac{T_{4c}}{M_2^2(N)}-1\nn \\
          &=& -\frac{32(N-4)(N^2-11N+36)}{N(N-1)(N-2)(N-3)}.
          \ee

          \subsection{The sixth order cumulant}

          In this subsection we evaluate the normalized sixth order cumulant which
          in terms of the moments is given by
          \be
          \kappa_6 = \frac{M_6(N)}{M_2^3(N)}- 15 \frac{M_4(N)}{M_2^2(N)}+30.
              \ee
           Since $M_4(N)$ was computed in the previous section we focus on $M_6(N)$.       
              The Gaussian integral for the sixth moment is again evaluated by summing
              over all pairwise contractions. In this case there are fifteen diagrams,
              and five of them are nested, see Fig. \ref{fig:dia} (6a-e). The nested diagrams are simply given
              by $M_2^3(N)$. The next simplest class of diagrams are those
              where two neighboring Hamiltonians are contracted, while
              the contractions of the remaining factors are intersecting,
              see Fig. \eref{fig:dia}(f-k). Their contribution to the
              sixth moment is given by
              \be
              T_{6f}=T_{6g}=T_{6h}=T_{6i}=T_{6j}=T_{6k} = M_2(N)T_{4c}.
              \ee

              By a cyclic permutation of the factors in $\Tr H^6$, it
              is clear that the diagrams in Fig. \ref{fig:dia} 6l-n are
              the same.
              If we fix the index of the second factor in diagram 6l,
              it is clear that by commuting  the factors as
              \be
              \Gamma_\alpha \Gamma_\beta \Gamma_\alpha \Gamma_\gamma \Gamma_\beta \Gamma_\gamma
              \to  \Gamma_\alpha \Gamma_\alpha \Gamma_\beta  \Gamma_\beta\Gamma_\gamma \Gamma_\gamma
              \ee
              we obtain the same combinatorial factor for the sum over $\alpha$ and $\gamma$ as in diagram
              4c. We thus find
              \be
              T_{6l}=T_{6m}=T_{6n}=           
              M_2^3(N){N \choose 4}^{-2}
   \left [\sum_{q=0}^4(-1)^q
{N-4 \choose 4-q}{4\choose q}
\right ]^2.
   \ee

   The most complicated diagram is diagram 6o corresponding to the trace
   \be
   Tr[\Gamma_\alpha \Gamma_\beta \Gamma_\gamma \Gamma_\alpha \Gamma_\beta \Gamma_\gamma].
   \ee
   The simplest way to do combinatorics is to think of $\Gamma_\beta \Gamma_\gamma$ as a
   product of 8 gamma matrices with $q$ gamma matrices in common while $\Gamma_\alpha $ share
   $l$ gamma matrices with   $ \Gamma_\beta \Gamma_\gamma$ and of those $l$ there are $l-m$ in the
   common factors. The result for this diagram is given by
\be
T_{6o}=          
M_2^3(N)
{N\choose 4}^{-2}
\sum_{q=0}^4 \sum_{l=0}^{4}\sum_{m=0}^{l}(-1)^{q+m}
    {N-8+q\choose 4-l} {8-2q \choose m}{N-4 \choose 4-q}{4\choose q}{q\choose l-m}.\nn\\
\ee
    
	Again, as a check of this result,  if the phase factor $(-1)^{q+m}$ is put to one, we find $M_2^3(d)$.
	
	Combining all contributions we find the normalized sixth cumulant
	\be
	\kappa_6(N)= \frac
	{512 (N-4) (11 N^5-304 N^4+3535 N^3-21302 N^2+65856 N-82656)}{(N-3)^2 (N-2)^2 (N-1)^2 N^2}.\nn\\
	\ee


\bibliography{library2}

\begin{thebibliography}{47}%
\makeatletter
\providecommand \@ifxundefined [1]{%
 \@ifx{#1\undefined}
}%
\providecommand \@ifnum [1]{%
 \ifnum #1\expandafter \@firstoftwo
 \else \expandafter \@secondoftwo
 \fi
}%
\providecommand \@ifx [1]{%
 \ifx #1\expandafter \@firstoftwo
 \else \expandafter \@secondoftwo
 \fi
}%
\providecommand \natexlab [1]{#1}%
\providecommand \enquote  [1]{``#1''}%
\providecommand \bibnamefont  [1]{#1}%
\providecommand \bibfnamefont [1]{#1}%
\providecommand \citenamefont [1]{#1}%
\providecommand \href@noop [0]{\@secondoftwo}%
\providecommand \href [0]{\begingroup \@sanitize@url \@href}%
\providecommand \@href[1]{\@@startlink{#1}\@@href}%
\providecommand \@@href[1]{\endgroup#1\@@endlink}%
\providecommand \@sanitize@url [0]{\catcode `\\12\catcode `\$12\catcode
  `\&12\catcode `\#12\catcode `\^12\catcode `\_12\catcode `\%12\relax}%
\providecommand \@@startlink[1]{}%
\providecommand \@@endlink[0]{}%
\providecommand \url  [0]{\begingroup\@sanitize@url \@url }%
\providecommand \@url [1]{\endgroup\@href {#1}{\urlprefix }}%
\providecommand \urlprefix  [0]{URL }%
\providecommand \Eprint [0]{\href }%
\providecommand \doibase [0]{http://dx.doi.org/}%
\providecommand \selectlanguage [0]{\@gobble}%
\providecommand \bibinfo  [0]{\@secondoftwo}%
\providecommand \bibfield  [0]{\@secondoftwo}%
\providecommand \translation [1]{[#1]}%
\providecommand \BibitemOpen [0]{}%
\providecommand \bibitemStop [0]{}%
\providecommand \bibitemNoStop [0]{.\EOS\space}%
\providecommand \EOS [0]{\spacefactor3000\relax}%
\providecommand \BibitemShut  [1]{\csname bibitem#1\endcsname}%
\let\auto@bib@innerbib\@empty
\bibitem [{\citenamefont {Wigner}(1951)}]{wigner1951}%
  \BibitemOpen
  \bibfield  {author} {\bibinfo {author} {\bibfnamefont {E.}~\bibnamefont
  {Wigner}},\ }\href {\doibase 10.1017/S0305004100027237} {\bibfield  {journal}
  {\bibinfo  {journal} {Math. Proc. Cam. Phil. Soc.}\ }\textbf {\bibinfo
  {volume} {49}},\ \bibinfo {pages} {790} (\bibinfo {year} {1951})}\BibitemShut
  {NoStop}%
\bibitem [{\citenamefont {Dyson}(1962{\natexlab{a}})}]{dyson1962a}%
  \BibitemOpen
  \bibfield  {author} {\bibinfo {author} {\bibfnamefont {F.}~\bibnamefont
  {Dyson}},\ }\href@noop {} {\bibfield  {journal} {\bibinfo  {journal} {J.
  Math. Phys.}\ }\textbf {\bibinfo {volume} {3}},\ \bibinfo {pages} {140}
  (\bibinfo {year} {1962}{\natexlab{a}})}\BibitemShut {NoStop}%
\bibitem [{\citenamefont {Dyson}(1962{\natexlab{b}})}]{dyson1962b}%
  \BibitemOpen
  \bibfield  {author} {\bibinfo {author} {\bibfnamefont {F.}~\bibnamefont
  {Dyson}},\ }\href@noop {} {\bibfield  {journal} {\bibinfo  {journal} {J.
  Math. Phys.}\ }\textbf {\bibinfo {volume} {3}},\ \bibinfo {pages} {157}
  (\bibinfo {year} {1962}{\natexlab{b}})}\BibitemShut {NoStop}%
\bibitem [{\citenamefont {Dyson}(1962{\natexlab{c}})}]{dyson1962c}%
  \BibitemOpen
  \bibfield  {author} {\bibinfo {author} {\bibfnamefont {F.}~\bibnamefont
  {Dyson}},\ }\href@noop {} {\bibfield  {journal} {\bibinfo  {journal} {J.
  Math. Phys.}\ }\textbf {\bibinfo {volume} {3}},\ \bibinfo {pages} {166}
  (\bibinfo {year} {1962}{\natexlab{c}})}\BibitemShut {NoStop}%
\bibitem [{\citenamefont {Dyson}(1962{\natexlab{d}})}]{dyson1962d}%
  \BibitemOpen
  \bibfield  {author} {\bibinfo {author} {\bibfnamefont {F.}~\bibnamefont
  {Dyson}},\ }\href@noop {} {\bibfield  {journal} {\bibinfo  {journal} {J.
  Math. Phys.}\ }\textbf {\bibinfo {volume} {3}},\ \bibinfo {pages} {1191}
  (\bibinfo {year} {1962}{\natexlab{d}})}\BibitemShut {NoStop}%
\bibitem [{\citenamefont {Dyson}(1972)}]{dyson1972}%
  \BibitemOpen
  \bibfield  {author} {\bibinfo {author} {\bibfnamefont {F.}~\bibnamefont
  {Dyson}},\ }\href@noop {} {\bibfield  {journal} {\bibinfo  {journal} {J.
  Math. Phys.}\ }\textbf {\bibinfo {volume} {13}},\ \bibinfo {pages} {90}
  (\bibinfo {year} {1972})}\BibitemShut {NoStop}%
\bibitem [{\citenamefont {Guhr}\ \emph {et~al.}(1998)\citenamefont {Guhr},
  \citenamefont {Mueller-Groeling},\ and\ \citenamefont
  {Weidenmueller}}]{guhr1998}%
  \BibitemOpen
  \bibfield  {author} {\bibinfo {author} {\bibfnamefont {T.}~\bibnamefont
  {Guhr}}, \bibinfo {author} {\bibfnamefont {A.}~\bibnamefont
  {Mueller-Groeling}}, \ and\ \bibinfo {author} {\bibfnamefont {H.~A.}\
  \bibnamefont {Weidenmueller}},\ }\href {\doibase
  http://dx.doi.org/10.1016/S0370-1573(97)00088-4} {\bibfield  {journal}
  {\bibinfo  {journal} {Physics Reports}\ }\textbf {\bibinfo {volume} {299}},\
  \bibinfo {pages} {189 } (\bibinfo {year} {1998})}\BibitemShut {NoStop}%
\bibitem [{\citenamefont {Bethe}(1936)}]{bethe1936}%
  \BibitemOpen
  \bibfield  {author} {\bibinfo {author} {\bibfnamefont {H.~A.}\ \bibnamefont
  {Bethe}},\ }\href {\doibase 10.1103/PhysRev.50.332} {\bibfield  {journal}
  {\bibinfo  {journal} {Phys. Rev.}\ }\textbf {\bibinfo {volume} {50}},\
  \bibinfo {pages} {332} (\bibinfo {year} {1936})}\BibitemShut {NoStop}%
\bibitem [{\citenamefont {Bohigas}\ and\ \citenamefont
  {Flores}(1971{\natexlab{a}})}]{bohigas1971}%
  \BibitemOpen
  \bibfield  {author} {\bibinfo {author} {\bibfnamefont {O.}~\bibnamefont
  {Bohigas}}\ and\ \bibinfo {author} {\bibfnamefont {J.}~\bibnamefont
  {Flores}},\ }\href {\doibase http://dx.doi.org/10.1016/0370-2693(71)90598-3}
  {\bibfield  {journal} {\bibinfo  {journal} {Physics Letters B}\ }\textbf
  {\bibinfo {volume} {34}},\ \bibinfo {pages} {261 } (\bibinfo {year}
  {1971}{\natexlab{a}})}\BibitemShut {NoStop}%
\bibitem [{\citenamefont {Bohigas}\ and\ \citenamefont
  {Flores}(1971{\natexlab{b}})}]{bohigas1971a}%
  \BibitemOpen
  \bibfield  {author} {\bibinfo {author} {\bibfnamefont {O.}~\bibnamefont
  {Bohigas}}\ and\ \bibinfo {author} {\bibfnamefont {J.}~\bibnamefont
  {Flores}},\ }\href {\doibase http://dx.doi.org/10.1016/0370-2693(71)90399-6}
  {\bibfield  {journal} {\bibinfo  {journal} {Physics Letters B}\ }\textbf
  {\bibinfo {volume} {35}},\ \bibinfo {pages} {383 } (\bibinfo {year}
  {1971}{\natexlab{b}})}\BibitemShut {NoStop}%
\bibitem [{\citenamefont {French}\ and\ \citenamefont
  {Wong}(1970)}]{french1970}%
  \BibitemOpen
  \bibfield  {author} {\bibinfo {author} {\bibfnamefont {J.}~\bibnamefont
  {French}}\ and\ \bibinfo {author} {\bibfnamefont {S.}~\bibnamefont {Wong}},\
  }\href {\doibase http://dx.doi.org/10.1016/0370-2693(70)90213-3} {\bibfield
  {journal} {\bibinfo  {journal} {Physics Letters B}\ }\textbf {\bibinfo
  {volume} {33}},\ \bibinfo {pages} {449 } (\bibinfo {year}
  {1970})}\BibitemShut {NoStop}%
\bibitem [{\citenamefont {French}\ and\ \citenamefont
  {Wong}(1971)}]{french1971}%
  \BibitemOpen
  \bibfield  {author} {\bibinfo {author} {\bibfnamefont {J.}~\bibnamefont
  {French}}\ and\ \bibinfo {author} {\bibfnamefont {S.}~\bibnamefont {Wong}},\
  }\href {\doibase http://dx.doi.org/10.1016/0370-2693(71)90424-2} {\bibfield
  {journal} {\bibinfo  {journal} {Physics Letters B}\ }\textbf {\bibinfo
  {volume} {35}},\ \bibinfo {pages} {5 } (\bibinfo {year} {1971})}\BibitemShut
  {NoStop}%
\bibitem [{\citenamefont {Mon}\ and\ \citenamefont {French}(1975)}]{mon1975}%
  \BibitemOpen
  \bibfield  {author} {\bibinfo {author} {\bibfnamefont {K.}~\bibnamefont
  {Mon}}\ and\ \bibinfo {author} {\bibfnamefont {J.}~\bibnamefont {French}},\
  }\href {\doibase http://dx.doi.org/10.1016/0003-4916(75)90045-7} {\bibfield
  {journal} {\bibinfo  {journal} {Annals of Physics}\ }\textbf {\bibinfo
  {volume} {95}},\ \bibinfo {pages} {90 } (\bibinfo {year} {1975})}\BibitemShut
  {NoStop}%
\bibitem [{\citenamefont {Verbaarschot}\ and\ \citenamefont
  {Zirnbauer}(1984)}]{verbaarschot1984}%
  \BibitemOpen
  \bibfield  {author} {\bibinfo {author} {\bibfnamefont {J.}~\bibnamefont
  {Verbaarschot}}\ and\ \bibinfo {author} {\bibfnamefont {M.}~\bibnamefont
  {Zirnbauer}},\ }\href {\doibase
  http://dx.doi.org/10.1016/0003-4916(84)90240-9} {\bibfield  {journal}
  {\bibinfo  {journal} {Annals of Physics}\ }\textbf {\bibinfo {volume}
  {158}},\ \bibinfo {pages} {78 } (\bibinfo {year} {1984})}\BibitemShut
  {NoStop}%
\bibitem [{\citenamefont {Benet}\ and\ \citenamefont
  {Weidenmüller}(2003)}]{benet2003}%
  \BibitemOpen
  \bibfield  {author} {\bibinfo {author} {\bibfnamefont {L.}~\bibnamefont
  {Benet}}\ and\ \bibinfo {author} {\bibfnamefont {H.~A.}\ \bibnamefont
  {Weidenmüller}},\ }\href {http://stacks.iop.org/0305-4470/36/i=12/a=340}
  {\bibfield  {journal} {\bibinfo  {journal} {Journal of Physics A:
  Mathematical and General}\ }\textbf {\bibinfo {volume} {36}},\ \bibinfo
  {pages} {3569} (\bibinfo {year} {2003})}\BibitemShut {NoStop}%
\bibitem [{\citenamefont {Gomez}\ \emph {et~al.}(2011)\citenamefont {Gomez},
  \citenamefont {Kar}, \citenamefont {Kota}, \citenamefont {Molina},
  \citenamefont {Relano},\ and\ \citenamefont {Retamosa}}]{gomez2011}%
  \BibitemOpen
  \bibfield  {author} {\bibinfo {author} {\bibfnamefont {J.}~\bibnamefont
  {Gomez}}, \bibinfo {author} {\bibfnamefont {K.}~\bibnamefont {Kar}}, \bibinfo
  {author} {\bibfnamefont {V.}~\bibnamefont {Kota}}, \bibinfo {author}
  {\bibfnamefont {R.}~\bibnamefont {Molina}}, \bibinfo {author} {\bibfnamefont
  {A.}~\bibnamefont {Relano}}, \ and\ \bibinfo {author} {\bibfnamefont
  {J.}~\bibnamefont {Retamosa}},\ }\href {\doibase
  http://dx.doi.org/10.1016/j.physrep.2010.11.003} {\bibfield  {journal}
  {\bibinfo  {journal} {Physics Reports}\ }\textbf {\bibinfo {volume} {499}},\
  \bibinfo {pages} {103 } (\bibinfo {year} {2011})}\BibitemShut {NoStop}%
\bibitem [{\citenamefont {Brody}\ \emph {et~al.}(1981)\citenamefont {Brody},
  \citenamefont {Flores}, \citenamefont {French}, \citenamefont {Mello},
  \citenamefont {Pandey},\ and\ \citenamefont {Wong}}]{brody1981}%
  \BibitemOpen
  \bibfield  {author} {\bibinfo {author} {\bibfnamefont {T.~A.}\ \bibnamefont
  {Brody}}, \bibinfo {author} {\bibfnamefont {J.}~\bibnamefont {Flores}},
  \bibinfo {author} {\bibfnamefont {J.~B.}\ \bibnamefont {French}}, \bibinfo
  {author} {\bibfnamefont {P.~A.}\ \bibnamefont {Mello}}, \bibinfo {author}
  {\bibfnamefont {A.}~\bibnamefont {Pandey}}, \ and\ \bibinfo {author}
  {\bibfnamefont {S.~S.~M.}\ \bibnamefont {Wong}},\ }\href {\doibase
  10.1103/RevModPhys.53.385} {\bibfield  {journal} {\bibinfo  {journal} {Rev.
  Mod. Phys.}\ }\textbf {\bibinfo {volume} {53}},\ \bibinfo {pages} {385}
  (\bibinfo {year} {1981})}\BibitemShut {NoStop}%
\bibitem [{\citenamefont {Kota}(2014)}]{kota2014}%
  \BibitemOpen
  \bibfield  {author} {\bibinfo {author} {\bibfnamefont {V.~K.~B.}\
  \bibnamefont {Kota}},\ }\href@noop {} {\emph {\bibinfo {title} {Embedded
  random matrix ensembles in quantum physics}}},\ Vol.\ \bibinfo {volume}
  {884}\ (\bibinfo  {publisher} {Springer},\ \bibinfo {year}
  {2014})\BibitemShut {NoStop}%
\bibitem [{\citenamefont {Kota}\ \emph {et~al.}(2011)\citenamefont {Kota},
  \citenamefont {Relano}, \citenamefont {Retamosa},\ and\ \citenamefont
  {Vyas}}]{kota2011a}%
  \BibitemOpen
  \bibfield  {author} {\bibinfo {author} {\bibfnamefont {V.~K.~B.}\
  \bibnamefont {Kota}}, \bibinfo {author} {\bibfnamefont {A.}~\bibnamefont
  {Relano}}, \bibinfo {author} {\bibfnamefont {J.}~\bibnamefont {Retamosa}}, \
  and\ \bibinfo {author} {\bibfnamefont {M.}~\bibnamefont {Vyas}},\ }\href
  {http://stacks.iop.org/1742-5468/2011/i=10/a=P10028} {\bibfield  {journal}
  {\bibinfo  {journal} {Journal of Statistical Mechanics: Theory and
  Experiment}\ }\textbf {\bibinfo {volume} {2011}},\ \bibinfo {pages} {P10028}
  (\bibinfo {year} {2011})}\BibitemShut {NoStop}%
\bibitem [{\citenamefont {Kitaev}()}]{kitaev2015}%
  \BibitemOpen
  \bibfield  {author} {\bibinfo {author} {\bibfnamefont {A.}~\bibnamefont
  {Kitaev}},\ }\href@noop {} {\enquote {\bibinfo {title} {A simple model of
  quantum holography},}\ }\bibinfo {note} {KITP strings seminar and
  Entanglement 2015 program, 12 February, 7 April and 27 May 2015,
  http://online.kitp.ucsb.edu/online/entangled15/}\BibitemShut {NoStop}%
\bibitem [{\citenamefont {Maldacena}\ and\ \citenamefont
  {Stanford}(2016)}]{maldacena2016}%
  \BibitemOpen
  \bibfield  {author} {\bibinfo {author} {\bibfnamefont {J.}~\bibnamefont
  {Maldacena}}\ and\ \bibinfo {author} {\bibfnamefont {D.}~\bibnamefont
  {Stanford}},\ }\href@noop {} {\bibfield  {journal} {\bibinfo  {journal}
  {arXiv preprint arXiv:1604.07818}\ } (\bibinfo {year} {2016})}\BibitemShut
  {NoStop}%
\bibitem [{\citenamefont {Polchinski}\ and\ \citenamefont
  {Rosenhaus}(2016)}]{polchinski2016}%
  \BibitemOpen
  \bibfield  {author} {\bibinfo {author} {\bibfnamefont {J.}~\bibnamefont
  {Polchinski}}\ and\ \bibinfo {author} {\bibfnamefont {V.}~\bibnamefont
  {Rosenhaus}},\ }\href {\doibase 10.1007/JHEP04(2016)001} {\bibfield
  {journal} {\bibinfo  {journal} {Journal of High Energy Physics}\ }\textbf
  {\bibinfo {volume} {2016}},\ \bibinfo {pages} {1} (\bibinfo {year}
  {2016})}\BibitemShut {NoStop}%
\bibitem [{\citenamefont {Engels{\"o}y}\ \emph {et~al.}(2016)\citenamefont
  {Engels{\"o}y}, \citenamefont {Mertens},\ and\ \citenamefont
  {Verlinde}}]{engels2016}%
  \BibitemOpen
  \bibfield  {author} {\bibinfo {author} {\bibfnamefont {J.}~\bibnamefont
  {Engels{\"o}y}}, \bibinfo {author} {\bibfnamefont {T.~G.}\ \bibnamefont
  {Mertens}}, \ and\ \bibinfo {author} {\bibfnamefont {H.}~\bibnamefont
  {Verlinde}},\ }\href {\doibase 10.1007/JHEP07(2016)139} {\bibfield  {journal}
  {\bibinfo  {journal} {Journal of High Energy Physics}\ }\textbf {\bibinfo
  {volume} {2016}},\ \bibinfo {pages} {1} (\bibinfo {year} {2016})}\BibitemShut
  {NoStop}%
\bibitem [{\citenamefont {Almheiri}\ and\ \citenamefont
  {Polchinski}(2015)}]{almheiri2015}%
  \BibitemOpen
  \bibfield  {author} {\bibinfo {author} {\bibfnamefont {A.}~\bibnamefont
  {Almheiri}}\ and\ \bibinfo {author} {\bibfnamefont {J.}~\bibnamefont
  {Polchinski}},\ }\href {\doibase 10.1007/JHEP11(2015)014} {\bibfield
  {journal} {\bibinfo  {journal} {Journal of High Energy Physics}\ }\textbf
  {\bibinfo {volume} {2015}},\ \bibinfo {pages} {1} (\bibinfo {year}
  {2015})}\BibitemShut {NoStop}%
\bibitem [{\citenamefont {Mag\'an}(2016)}]{magan2016}%
  \BibitemOpen
  \bibfield  {author} {\bibinfo {author} {\bibfnamefont {J.~M.}\ \bibnamefont
  {Mag\'an}},\ }\href {\doibase 10.1103/PhysRevLett.116.030401} {\bibfield
  {journal} {\bibinfo  {journal} {Phys. Rev. Lett.}\ }\textbf {\bibinfo
  {volume} {116}},\ \bibinfo {pages} {030401} (\bibinfo {year}
  {2016})}\BibitemShut {NoStop}%
\bibitem [{\citenamefont {Danshita}\ \emph {et~al.}(2016)\citenamefont
  {Danshita}, \citenamefont {Hanada},\ and\ \citenamefont
  {Tezuka}}]{danshita2016}%
  \BibitemOpen
  \bibfield  {author} {\bibinfo {author} {\bibfnamefont {I.}~\bibnamefont
  {Danshita}}, \bibinfo {author} {\bibfnamefont {M.}~\bibnamefont {Hanada}}, \
  and\ \bibinfo {author} {\bibfnamefont {M.}~\bibnamefont {Tezuka}},\
  }\href@noop {} {\bibfield  {journal} {\bibinfo  {journal} {arXiv preprint
  arXiv:1606.02454}\ } (\bibinfo {year} {2016})}\BibitemShut {NoStop}%
\bibitem [{\citenamefont {Garcia-Alvarez}\ \emph {et~al.}(2016)\citenamefont
  {Garcia-Alvarez}, \citenamefont {Egusquiza}, \citenamefont {Lamata},
  \citenamefont {del Campo}, \citenamefont {Sonner},\ and\ \citenamefont
  {Solano}}]{garcia-alvarez2016}%
  \BibitemOpen
  \bibfield  {author} {\bibinfo {author} {\bibfnamefont {L.}~\bibnamefont
  {Garcia-Alvarez}}, \bibinfo {author} {\bibfnamefont {I.~L.}\ \bibnamefont
  {Egusquiza}}, \bibinfo {author} {\bibfnamefont {L.}~\bibnamefont {Lamata}},
  \bibinfo {author} {\bibfnamefont {A.}~\bibnamefont {del Campo}}, \bibinfo
  {author} {\bibfnamefont {J.}~\bibnamefont {Sonner}}, \ and\ \bibinfo {author}
  {\bibfnamefont {E.}~\bibnamefont {Solano}},\ }\href@noop {} {\  (\bibinfo
  {year} {2016})},\ \Eprint {http://arxiv.org/abs/1607.08560} {arXiv:1607.08560
  [quant-ph]} \BibitemShut {NoStop}%
\bibitem [{\citenamefont {Bagrets}\ \emph {et~al.}(2016)\citenamefont
  {Bagrets}, \citenamefont {Altland},\ and\ \citenamefont
  {Kamenev}}]{bagrets2016}%
  \BibitemOpen
  \bibfield  {author} {\bibinfo {author} {\bibfnamefont {D.}~\bibnamefont
  {Bagrets}}, \bibinfo {author} {\bibfnamefont {A.}~\bibnamefont {Altland}}, \
  and\ \bibinfo {author} {\bibfnamefont {A.}~\bibnamefont {Kamenev}},\
  }\href@noop {} {\bibfield  {journal} {\bibinfo  {journal} {Nuclear Physics
  B}\ }\textbf {\bibinfo {volume} {911}},\ \bibinfo {pages} {191} (\bibinfo
  {year} {2016})}\BibitemShut {NoStop}%
\bibitem [{\citenamefont {Sachdev}(2015)}]{sachdev2015}%
  \BibitemOpen
  \bibfield  {author} {\bibinfo {author} {\bibfnamefont {S.}~\bibnamefont
  {Sachdev}},\ }\href {\doibase 10.1103/PhysRevX.5.041025} {\bibfield
  {journal} {\bibinfo  {journal} {Phys. Rev. X}\ }\textbf {\bibinfo {volume}
  {5}},\ \bibinfo {pages} {041025} (\bibinfo {year} {2015})}\BibitemShut
  {NoStop}%
\bibitem [{\citenamefont {You}\ \emph {et~al.}(2016)\citenamefont {You},
  \citenamefont {Ludwig},\ and\ \citenamefont {Xu}}]{you2016}%
  \BibitemOpen
  \bibfield  {author} {\bibinfo {author} {\bibfnamefont {Y.-Z.}\ \bibnamefont
  {You}}, \bibinfo {author} {\bibfnamefont {A.~W.}\ \bibnamefont {Ludwig}}, \
  and\ \bibinfo {author} {\bibfnamefont {C.}~\bibnamefont {Xu}},\ }\href@noop
  {} {\bibfield  {journal} {\bibinfo  {journal} {arXiv preprint
  arXiv:1602.06964}\ } (\bibinfo {year} {2016})}\BibitemShut {NoStop}%
\bibitem [{\citenamefont {Gross}\ and\ \citenamefont
  {Rosenhaus}(2016)}]{Gross:2016kjj}%
  \BibitemOpen
  \bibfield  {author} {\bibinfo {author} {\bibfnamefont {D.~J.}\ \bibnamefont
  {Gross}}\ and\ \bibinfo {author} {\bibfnamefont {V.}~\bibnamefont
  {Rosenhaus}},\ }\href@noop {} {\  (\bibinfo {year} {2016})},\ \Eprint
  {http://arxiv.org/abs/1610.01569} {arXiv:1610.01569 [hep-th]} \BibitemShut
  {NoStop}%
\bibitem [{\citenamefont {Sachdev}\ and\ \citenamefont
  {Ye}(1993)}]{sachdev1993}%
  \BibitemOpen
  \bibfield  {author} {\bibinfo {author} {\bibfnamefont {S.}~\bibnamefont
  {Sachdev}}\ and\ \bibinfo {author} {\bibfnamefont {J.}~\bibnamefont {Ye}},\
  }\href {\doibase 10.1103/PhysRevLett.70.3339} {\bibfield  {journal} {\bibinfo
   {journal} {Phys. Rev. Lett.}\ }\textbf {\bibinfo {volume} {70}},\ \bibinfo
  {pages} {3339} (\bibinfo {year} {1993})}\BibitemShut {NoStop}%
\bibitem [{\citenamefont {Maldacena}(1999)}]{maldacena1998}%
  \BibitemOpen
  \bibfield  {author} {\bibinfo {author} {\bibfnamefont {J.}~\bibnamefont
  {Maldacena}},\ }\href@noop {} {\bibfield  {journal} {\bibinfo  {journal}
  {International journal of theoretical physics}\ }\textbf {\bibinfo {volume}
  {38}},\ \bibinfo {pages} {1113} (\bibinfo {year} {1999})}\BibitemShut
  {NoStop}%
\bibitem [{\citenamefont {Jensen}(2016)}]{jensen2016}%
  \BibitemOpen
  \bibfield  {author} {\bibinfo {author} {\bibfnamefont {K.}~\bibnamefont
  {Jensen}},\ }\href {\doibase 10.1103/PhysRevLett.117.111601} {\bibfield
  {journal} {\bibinfo  {journal} {Phys. Rev. Lett.}\ }\textbf {\bibinfo
  {volume} {117}},\ \bibinfo {pages} {111601} (\bibinfo {year}
  {2016})}\BibitemShut {NoStop}%
\bibitem [{\citenamefont {Cvetič}\ and\ \citenamefont
  {Papadimitriou}(2016)}]{cvetic2016}%
  \BibitemOpen
  \bibfield  {author} {\bibinfo {author} {\bibfnamefont {M.}~\bibnamefont
  {Cvetič}}\ and\ \bibinfo {author} {\bibfnamefont {I.}~\bibnamefont
  {Papadimitriou}},\ }\href@noop {} {\  (\bibinfo {year} {2016})},\ \Eprint
  {http://arxiv.org/abs/1608.07018} {arXiv:1608.07018 [hep-th]} \BibitemShut
  {NoStop}%
\bibitem [{\citenamefont {Jevicki}\ \emph {et~al.}(2016)\citenamefont
  {Jevicki}, \citenamefont {Suzuki},\ and\ \citenamefont {Yoon}}]{jevicki2016}%
  \BibitemOpen
  \bibfield  {author} {\bibinfo {author} {\bibfnamefont {A.}~\bibnamefont
  {Jevicki}}, \bibinfo {author} {\bibfnamefont {K.}~\bibnamefont {Suzuki}}, \
  and\ \bibinfo {author} {\bibfnamefont {J.}~\bibnamefont {Yoon}},\ }\href
  {\doibase 10.1007/JHEP07(2016)007} {\bibfield  {journal} {\bibinfo  {journal}
  {Journal of High Energy Physics}\ }\textbf {\bibinfo {volume} {2016}},\
  \bibinfo {pages} {1} (\bibinfo {year} {2016})}\BibitemShut {NoStop}%
\bibitem [{\citenamefont {Maldacena}\ \emph {et~al.}(2015)\citenamefont
  {Maldacena}, \citenamefont {Shenker},\ and\ \citenamefont
  {Stanford}}]{maldacena2015}%
  \BibitemOpen
  \bibfield  {author} {\bibinfo {author} {\bibfnamefont {J.}~\bibnamefont
  {Maldacena}}, \bibinfo {author} {\bibfnamefont {S.~H.}\ \bibnamefont
  {Shenker}}, \ and\ \bibinfo {author} {\bibfnamefont {D.}~\bibnamefont
  {Stanford}},\ }\href@noop {} {\bibfield  {journal} {\bibinfo  {journal}
  {arXiv preprint arXiv:1503.01409}\ } (\bibinfo {year} {2015})}\BibitemShut
  {NoStop}%
\bibitem [{\citenamefont {Witten}(2016)}]{witten2016}%
  \BibitemOpen
  \bibfield  {author} {\bibinfo {author} {\bibfnamefont {E.}~\bibnamefont
  {Witten}},\ }\href@noop {} {\  (\bibinfo {year} {2016})},\ \Eprint
  {http://arxiv.org/abs/1610.09758} {arXiv:1610.09758 [hep-th]} \BibitemShut
  {NoStop}%
\bibitem [{\citenamefont {Altshuler}\ \emph {et~al.}(1988)\citenamefont
  {Altshuler}, \citenamefont {Zarekeshev}, \citenamefont {Kotochigova},\ and\
  \citenamefont {Shklovskii}}]{altshuler1988}%
  \BibitemOpen
  \bibfield  {author} {\bibinfo {author} {\bibfnamefont {B.}~\bibnamefont
  {Altshuler}}, \bibinfo {author} {\bibfnamefont {I.}~\bibnamefont
  {Zarekeshev}}, \bibinfo {author} {\bibfnamefont {S.}~\bibnamefont
  {Kotochigova}}, \ and\ \bibinfo {author} {\bibfnamefont {B.}~\bibnamefont
  {Shklovskii}},\ }\href@noop {} {\bibfield  {journal} {\bibinfo  {journal}
  {Sov. Phys. JETP [Zh. Eksp. Teor. Fiz. 94, 343]}\ }\textbf {\bibinfo {volume}
  {67}},\ \bibinfo {pages} {15} (\bibinfo {year} {1988})}\BibitemShut {NoStop}%
\bibitem [{\citenamefont {Braun}\ and\ \citenamefont
  {Montambaux}(1995)}]{braun1995}%
  \BibitemOpen
  \bibfield  {author} {\bibinfo {author} {\bibfnamefont {D.}~\bibnamefont
  {Braun}}\ and\ \bibinfo {author} {\bibfnamefont {G.}~\bibnamefont
  {Montambaux}},\ }\href {\doibase 10.1103/PhysRevB.52.13903} {\bibfield
  {journal} {\bibinfo  {journal} {Phys. Rev. B}\ }\textbf {\bibinfo {volume}
  {52}},\ \bibinfo {pages} {13903} (\bibinfo {year} {1995})}\BibitemShut
  {NoStop}%
\bibitem [{\citenamefont {Bertrand}\ and\ \citenamefont
  {Garc\'{\i}a-Garc\'{\i}a}(2016)}]{bertrand2016}%
  \BibitemOpen
  \bibfield  {author} {\bibinfo {author} {\bibfnamefont {C.~L.}\ \bibnamefont
  {Bertrand}}\ and\ \bibinfo {author} {\bibfnamefont {A.~M.}\ \bibnamefont
  {Garc\'{\i}a-Garc\'{\i}a}},\ }\href {\doibase 10.1103/PhysRevB.94.144201}
  {\bibfield  {journal} {\bibinfo  {journal} {Phys. Rev. B}\ }\textbf {\bibinfo
  {volume} {94}},\ \bibinfo {pages} {144201} (\bibinfo {year}
  {2016})}\BibitemShut {NoStop}%
\bibitem [{\citenamefont {Benet}\ \emph {et~al.}(2001)\citenamefont {Benet},
  \citenamefont {Rupp},\ and\ \citenamefont {Weidenm\"uller}}]{benet2001}%
  \BibitemOpen
  \bibfield  {author} {\bibinfo {author} {\bibfnamefont {L.}~\bibnamefont
  {Benet}}, \bibinfo {author} {\bibfnamefont {T.}~\bibnamefont {Rupp}}, \ and\
  \bibinfo {author} {\bibfnamefont {H.~A.}\ \bibnamefont {Weidenm\"uller}},\
  }\href {\doibase 10.1103/PhysRevLett.87.010601} {\bibfield  {journal}
  {\bibinfo  {journal} {Phys. Rev. Lett.}\ }\textbf {\bibinfo {volume} {87}},\
  \bibinfo {pages} {010601} (\bibinfo {year} {2001})}\BibitemShut {NoStop}%
\bibitem [{\citenamefont {Srednicki}(2002)}]{srednicki2002}%
  \BibitemOpen
  \bibfield  {author} {\bibinfo {author} {\bibfnamefont {M.}~\bibnamefont
  {Srednicki}},\ }\href {\doibase 10.1103/PhysRevE.66.046138} {\bibfield
  {journal} {\bibinfo  {journal} {Phys. Rev. E}\ }\textbf {\bibinfo {volume}
  {66}},\ \bibinfo {pages} {046138} (\bibinfo {year} {2002})}\BibitemShut
  {NoStop}%
\bibitem [{\citenamefont {Porter}(1965)}]{porter}%
  \BibitemOpen
  \bibfield  {author} {\bibinfo {author} {\bibfnamefont {C.~E.}\ \bibnamefont
  {Porter}},\ }\href@noop {} {\emph {\bibinfo {title} {Statistical Theory of
  Spectra: Fluctuations}}}\ (\bibinfo  {publisher} {Academic Press},\ \bibinfo
  {year} {1965})\BibitemShut {NoStop}%
\bibitem [{\citenamefont {Kieburg}\ \emph {et~al.}(2015)\citenamefont
  {Kieburg}, \citenamefont {Verbaarschot},\ and\ \citenamefont
  {Zafeiropoulos}}]{kieburg2015}%
  \BibitemOpen
  \bibfield  {author} {\bibinfo {author} {\bibfnamefont {M.}~\bibnamefont
  {Kieburg}}, \bibinfo {author} {\bibfnamefont {J.~J.~M.}\ \bibnamefont
  {Verbaarschot}}, \ and\ \bibinfo {author} {\bibfnamefont {S.}~\bibnamefont
  {Zafeiropoulos}},\ }\href {\doibase 10.1103/PhysRevD.92.045026} {\bibfield
  {journal} {\bibinfo  {journal} {Phys. Rev.}\ }\textbf {\bibinfo {volume}
  {D92}},\ \bibinfo {pages} {045026} (\bibinfo {year} {2015})},\ \Eprint
  {http://arxiv.org/abs/1505.01784} {arXiv:1505.01784 [hep-lat]} \BibitemShut
  {NoStop}%
\bibitem [{\citenamefont {Maldacena}\ \emph {et~al.}(2016)\citenamefont
  {Maldacena}, \citenamefont {Stanford},\ and\ \citenamefont
  {Yang}}]{maldacena2016a}%
  \BibitemOpen
  \bibfield  {author} {\bibinfo {author} {\bibfnamefont {J.}~\bibnamefont
  {Maldacena}}, \bibinfo {author} {\bibfnamefont {D.}~\bibnamefont {Stanford}},
  \ and\ \bibinfo {author} {\bibfnamefont {Z.}~\bibnamefont {Yang}},\
  }\href@noop {} {\bibfield  {journal} {\bibinfo  {journal} {arXiv preprint
  arXiv:1606.01857}\ } (\bibinfo {year} {2016})}\BibitemShut {NoStop}%
\bibitem [{\citenamefont {Cotler}\ \emph {et~al.}(2016)\citenamefont {Cotler},
  \citenamefont {Gur-Ari}, \citenamefont {Hanada}, \citenamefont {Polchinski},
  \citenamefont {Saad}, \citenamefont {Shenker}, \citenamefont {Stanford},
  \citenamefont {Streicher},\ and\ \citenamefont {Tezuka}}]{cotler2016}%
  \BibitemOpen
  \bibfield  {author} {\bibinfo {author} {\bibfnamefont {J.~S.}\ \bibnamefont
  {Cotler}}, \bibinfo {author} {\bibfnamefont {G.}~\bibnamefont {Gur-Ari}},
  \bibinfo {author} {\bibfnamefont {M.}~\bibnamefont {Hanada}}, \bibinfo
  {author} {\bibfnamefont {J.}~\bibnamefont {Polchinski}}, \bibinfo {author}
  {\bibfnamefont {P.}~\bibnamefont {Saad}}, \bibinfo {author} {\bibfnamefont
  {S.~H.}\ \bibnamefont {Shenker}}, \bibinfo {author} {\bibfnamefont
  {D.}~\bibnamefont {Stanford}}, \bibinfo {author} {\bibfnamefont
  {A.}~\bibnamefont {Streicher}}, \ and\ \bibinfo {author} {\bibfnamefont
  {M.}~\bibnamefont {Tezuka}},\ }\href@noop {} {\  (\bibinfo {year} {2016})},\
  \Eprint {http://arxiv.org/abs/1611.04650} {arXiv:1611.04650 [hep-th]}
  \BibitemShut {NoStop}%
\end{thebibliography}%

\end{document}